\documentclass[aps,prb,superscriptaddress,reprint]{revtex4-1}

\usepackage{graphicx}
\usepackage{subfigure}
\usepackage{docs}
\usepackage{bm}
\usepackage{amssymb}
\usepackage{amsmath}
\usepackage[colorlinks=true,linkcolor=blue]{hyperref}
\expandafter\ifx\csname package@font\endcsname\relax\else
 \expandafter\expandafter
 \expandafter\usepackage
 \expandafter\expandafter
 \expandafter{\csname package@font\endcsname}%
\fi
\begin{document}

\widetext
\title{Theory and design of In$_{x}$Ga$_{1-x}$As$_{1-y}$Bi$_{y}$ mid-infrared semiconductor lasers:\\type-I quantum wells for emission beyond 3 $\mu$m on InP substrates}


\author{Christopher A.~Broderick}
\email{c.broderick@umail.ucc.ie} 
\affiliation{Tyndall National Institute, Lee Maltings, Dyke Parade, Cork T12 R5CP, Ireland}
\affiliation{Department of Physics, University College Cork, Cork T12 YN60, Ireland}
\affiliation{Department of Electrical and Electronic Engineering, University of Bristol, Bristol BS8 1UB, U.K.}

\author{Wanshu Xiong}
\affiliation{Department of Electrical and Electronic Engineering, University of Bristol, Bristol BS8 1UB, U.K.}

\author{Stephen J.~Sweeney}
\affiliation{Advanced Technology Institute and Department of Physics, University of Surrey, Guildford GU2 7XH, U.K.}

\author{Eoin P.~O'Reilly}
\affiliation{Tyndall National Institute, Lee Maltings, Dyke Parade, Cork T12 R5CP, Ireland}
\affiliation{Department of Physics, University College Cork, Cork T12 YN60, Ireland}

\author{Judy M.~Rorison}
\affiliation{Department of Electrical and Electronic Engineering, University of Bristol, Bristol BS8 1UB, U.K.}

\date{\today}


\begin{abstract}

We present a theoretical analysis and optimisation of the properties and performance of mid-infrared semiconductor lasers based on the dilute bismide alloy In$_{x}$Ga$_{1-x}$As$_{1-y}$Bi$_{y}$, grown on conventional (001) InP substrates. The ability to independently vary the epitaxial strain and emission wavelength in this quaternary alloy provides significant scope for band structure engineering. Our calculations demonstrate that structures based on compressively strained In$_{x}$Ga$_{1-x}$As$_{1-y}$Bi$_{y}$ quantum wells (QWs) can readily achieve emission wavelengths in the 3 -- 5 $\mu$m range, and that these QWs have large type-I band offsets. As such, these structures have the potential to overcome a number of limitations commonly associated with this application-rich but technologically challenging wavelength range. By considering structures having (i) fixed QW thickness and variable strain, and (ii) fixed strain and variable QW thickness, we quantify key trends in the properties and performance as functions of the alloy composition, structural properties, and emission wavelength, and on this basis identify routes towards the realisation of optimised devices for practical applications. Our analysis suggests that simple laser structures -- incorporating In$_{x}$Ga$_{1-x}$As$_{1-y}$Bi$_{y}$ QWs and unstrained ternary In$_{0.53}$Ga$_{0.47}$As barriers -- which are compatible with established epitaxial growth, provide a route to realising InP-based mid-infrared diode lasers.

\end{abstract}


\maketitle


\section{Introduction}
\label{sec:introduction}

The vibrational-rotational spectra of many environmentally and industrially important gases are characterised by strong absorption features at mid-infrared wavelengths. As such, there is broad scope for sensing applications of semiconductor lasers and light-emitting diodes operating in the 2 -- 5 $\mu$m wavelength range, including: detection of environmental pollutants, chemical monitoring in industrial processes, remote analysis of hazardous substances, and detection of biological markers in non-invasive medical diagnostics, in addition to applications in free-space optical communications and infrared countermeasures. \cite{Krier_book_2006,Bauer_SST_2011,Hodgkinson_MST_2013} However, the mid-infrared spectral region is particularly challenging from the perspective of semiconductor laser design and fabrication. \cite{Crowley_PJ_2012} This has led to a proliferation of approaches aiming to realise compact, efficient and inexpensive devices. \cite{Tournie_SS_2012,Jung_JO_2017} In the 2 -- 3 $\mu$m wavelength range diode lasers based on type-I In$_{x}$Ga$_{1-x}$As$_{1-y}$Sb$_{y}$/GaSb quantum wells (QWs) have become well established. \cite{Kim_APL_2002,Vizbaras_EL_2011,Vizbaras_SST_2012,Sifferman_JSTQE_2015,Andrejew_OL_2016} While such devices have reasonable characteristics their performance degrades with increasing wavelength, with high threshold currents and high temperature sensitivity associated with non-radiative Auger recombination processes. \cite{Sifferman_JSTQE_2015,Eales_JSTQE_2017} Good performance has been obtained in the 3 -- 4 $\mu$m wavelength range using inter-band cascade lasers (ICLs). \cite{Vurgaftman_JSTQE_2013,Vurgaftman_JPDAP_2015} However, ICLs are highly complicated structures requiring sophisticated growth, and also suffer from losses related to Auger recombination. Finally, above 4 $\mu$m quantum cascade lasers (QCLs) dominate. \cite{Razeghi_OE_2015} QCLs are complicated structures, often requiring in excess of 100 layers with very tight tolerances on the layer thicknesses. \cite{Vitiello_OE_2015} The performance of such devices degrades towards shorter wavelengths, owing to carrier leakage from the upper electron levels which gives rise to a strongly temperature-dependent threshold current. Consequently, in order to achieve sufficient performance to facilitate practical applications, there remain significant challenges associated with the development of efficient and low-cost devices operating at wavelengths between 3 and 4 $\mu$m.

Given their design and fabrication simplicity compared to cascade devices, there is strong motivation to develop diode lasers operating in this wavelength range. Currently there are two primary approaches to achieve this aim, focused respectively on GaSb- and InP-based structures. GaSb-based devices incorporating type-I In$_{x}$Ga$_{1-x}$As$_{1-y}$Sb$_{y}$ QWs have demonstrated impressive characteristics at emission wavelengths between 2 and 3 $\mu$m. \cite{Kim_APL_2002,Andrejew_OL_2016} Recently, room temperature operation of In$_{x}$Ga$_{1-x}$As$_{1-y}$Sb$_{y}$ devices out to wavelengths of 3.4 and 3.7 $\mu$m under continuous-wave and pulsed-mode operation, respectively, has been demonstrated. \cite{Hosoda_EL_2010,Vizbaras_SST_2012} However, these devices suffer from strong temperature dependence of the threshold current density -- which is exacerbated at longer wavelengths -- resulting from a combination of inter-valence band absorption (IVBA), non-radiative Auger recombination, and thermal carrier leakage. \cite{Brien_APL_2006,Tossou_SST_2013,Eales_JSTQE_2017} In parallel, significant effort has been dedicated to extending the wavelength range accessible using InP-based devices. \cite{Gu_chapter_2015} This has involved the development of a variety of novel heterostructures, based on (i) type-II structures incorporating In$_{x}$Ga$_{1-x}$(N)As/GaAs$_{y}$Sb$_{1-y}$ QWs (where pulsed-mode operation has been demonstrated out to 2.6 $\mu$m at 270 K), \cite{Vurgaftman_JAP_2004,Mawst_JSTQE_2008} or (ii) combining growth on Al$_{x}$In$_{1-x}$As or InAs$_{1-x}$P$_{x}$ metamorphic buffer layers with highly-strained In$_{x}$Ga$_{1-x}$As, InAs$_{1-x}$Sb$_{x}$, GaSb$_{1-x}$Bi$_{x}$ or InAs$_{1-x}$Bi$_{x}$ type-I QWs (where pulsed-mode operation of a 2.9 $\mu$m device at 230 K has recently been demonstrated). \cite{Jung_APL_2012,Cao_APL_2013,Gu_JPDAP_2013,Gu_APE_2014,Jung_APL_2015,Gu_APL_2016,Delorme_APL_2017} Despite these ongoing innovations in heterostructure design and fabrication, pushing the emission wavelength of InP-based diode lasers beyond 3 $\mu$m remains a significant challenge.

From a practical perspective is it desirable to develop InP- rather than GaSb-based devices, for several reasons. Firstly, InP substrates are significantly cheaper than their GaSb counterparts and, as a mainstay of the photonics industry, benefit from advanced fabrication and processing techniques. Secondly, the higher thermal conductivity of InP enhances heat dissipation, thereby allowing for more effective thermal management during operation. Thirdly, the InP platform is more technologically advanced, presenting the opportunity to take advantage of the capabilities offered by InP-based passive optical components and photonic integrated circuitry.

Here, we propose that InP-based mid-infrared diode lasers can be achieved using type-I QWs based on the dilute bismide alloy In$_{x}$Ga$_{1-x}$As$_{1-y}$Bi$_{y}$ (containing bismuth, Bi). \cite{Jin_JAP_2013,Gladysiewicz_JAP_2015,Broderick_NUSOD_2016_InP} We perform a theoretical analysis that identifies several properties of these structures which have the potential to overcome key limitations associated with existing GaSb-based devices operating in the 3 -- 4 $\mu$m range, while simultaneously enabling emission beyond 3 $\mu$m from an InP-based diode structure. The prototypical laser structures we consider -- which are chosen to be fully compatible with established InP-based growth and fabrication, as well as the epitaxial growth of In$_{x}$Ga$_{1-x}$As$_{1-y}$Bi$_{y}$ alloys -- can be grown pseudomorphically on InP substrates, removing the requirement to employ metamorphic buffer layers to facilitate the growth of lattice-mismatched heterostructures. We present an analysis of the In$_{x}$Ga$_{1-x}$As$_{1-y}$Bi$_{y}$ bulk band structure, highlighting the flexibility with which it can be engineered to achieve favourable characteristics for laser applications. In particular, we show that it is possible to engineer a band structure in which the valence band (VB) spin-orbit splitting energy $\Delta_{\scalebox{0.7}{\textrm{SO}}}$ exceeds the band gap $E_{g}$ -- as has been demonstrated for bulk alloys -- and that this can be achieved at significantly lower Bi compositions than in ternary GaAs$_{1-y}$Bi$_{y}$ alloys. \cite{Broderick_PSSB_2013,Jin_JAP_2013,Chai_SST_2015} This band structure condition is particularly favourable for light-emitting devices, since it is expected to reduce losses related to IVBA and non-radiative (CHSH) Auger recombination processes involving the spin-split-off (SO) VB, and is hence expected to contribute to increased device efficiency and thermal stability. \cite{Sweeney_IEEEISLC_2010,Sweeney_ICTON_2011,Broderick_SST_2012,Sweeney_JAP_2013,Broderick_bismide_chapter_2017}

We undertake a systematic investigation of the properties of prototypical In$_{x}$Ga$_{1-x}$As$_{1-y}$Bi$_{y}$ QW laser structures designed to emit between 3 and 5 $\mu$m. By considering structures having (i) fixed QW thickness and variable strain, and (ii) fixed strain and variable QW thickness, we quantify key trends in the expected laser performance as functions of the alloy composition, structural properties, and emission wavelength. Our results highlight the importance of epitaxial strain as the primary factor influencing performance, placing strong constraints on the ranges of In and Bi compositions of interest for the growth of QWs for practical devices. These QWs have several attractive properties, including large type-I band offsets -- to provide strong carrier confinement and suppressed thermal carrier leakage -- as well as high material gain, low threshold carrier densities and high differential gain. Overall, we identify and quantify pathways towards the design and realisation of optimised devices, and highlight that pseudomorphic In$_{x}$Ga$_{1-x}$As$_{1-y}$Bi$_{y}$ QW structures grown on conventional (001) InP substrates are particularly promising for the development of diode lasers operating in the 3 -- 4 $\mu$m wavelength range.

The remainder of this paper is organised as follows. In Sec.~\ref{sec:qw_properties} we outline the theoretical model we have developed to calculate the electronic and optical properties of In$_{x}$Ga$_{1-x}$As$_{1-y}$Bi$_{y}$ QWs, and in Sec.~\ref{sec:laser_modelling} provide a description of the laser structures on which our analysis focuses. Our results are presented in Sec.~\ref{sec:results}, beginning in Sec.~\ref{sec:InGaAsBi_band_structure} with a description of the potential for band structure engineering in quaternary In$_{x}$Ga$_{1-x}$As$_{1-y}$Bi$_{y}$ alloys. In Sec.~\ref{sec:3500_nm_structures} we undertake an analysis and optimisation of single QW laser structures designed to emit at 3.5 $\mu$m, before considering in Sec.~\ref{sec:longer_wavelength} the dependence of the laser characteristics on the emission wavelength and number of QWs throughout the 3 -- 5 $\mu$m range. Finally, in Sec.~\ref{sec:conclusions}, we summarise and conclude.


\section{Theoretical model}
\label{sec:theoretical_model}

Our theoretical model is closely based upon that we have developed previously to analyse the electronic and optical properties of GaAs$_{1-x}$Bi$_{x}$/(Al)GaAs QWs. \cite{Broderick_JSTQE_2015} Our previous work has demonstrated that this model is capable of quantitatively predicting the measured gain characteristics of real GaAs$_{1-x}$Bi$_{x}$-based QW lasers. \cite{Marko_SR_2016} This analysis has underscored the importance of accurately accounting for the Bi-induced modifications to the band structure in calculations of the electronic and optical properties. Here, we outline the extension of this model to treat quaternary In$_{x}$Ga$_{1-x}$As$_{1-y}$Bi$_{y}$ QWs. A comprehensive description of the details of our theoretical approach can be found in our recently published review of the theory and simulation of dilute bismide alloys, Ref.~\onlinecite{Broderick_bismide_chapter_2017}.


\subsection{QW electronic and optical properties}
\label{sec:qw_properties}

When substituted in place of As in In$_{x}$Ga$_{1-x}$As Bi acts as an isovalent impurity, giving rise to a set of degenerate Bi-derived localised impurity states which are resonant with the VB states of the In$_{x}$Ga$_{1-x}$As host matrix semiconductor. \cite{Zhang_PRB_2005,Usman_PRB_2011} These Bi-related localised states couple to the extended In$_{x}$Ga$_{1-x}$As host matrix VB edge states via a Bi composition-dependent valence band-anticrossing (VBAC) interaction. \cite{Alberi_PRB_2007,Usman_PRB_2011} As a result of this interaction, the In$_{x}$Ga$_{1-x}$As$_{1-y}$Bi$_{y}$ alloy VB edge states are then formed of an admixture of the Bi-related localised and extended host matrix VB edge states. We have previously demonstrated that this unusual material behaviour can be described quantitatively using an extended basis 12-band \textbf{k}$\cdot$\textbf{p} Hamiltonian, in which the conventional 8-band basis set -- consisting of the zone-centre Bloch functions associated with the lowest energy conduction band (CB), as well as the light-hole (LH), heavy-hole (HH), and spin-split-off (SO) VBs \cite{Bahder_PRB_1990} -- is augmented by the inclusion of a set of four degenerate Bi-related localised states, which can be chosen to have LH- and HH-like symmetry. \cite{Broderick_PSSB_2013,Broderick_SST_2013}

Via detailed theoretical analysis of the electronic structure of (In,Ga)As$_{1-y}$Bi$_{y}$ alloys we have found that the impact of Bi incorporation on the band structure can be described via the 12-band Hamiltonian using seven independent Bi-related parameters: (i) $\Delta E_{\scalebox{0.7}{\textrm{Bi}}}$, the energy of the Bi-related localised states relative to the unperturbed host matrix VB edge, (ii) $\beta$, which determines the magnitude $\beta \sqrt{y}$ of the Bi composition dependent VBAC matrix elements coupling the Bi-related localised and extended host matrix LH- and HH-like VB edge states, (iii) -- (v) $\alpha$, $\kappa$ and $\gamma$, which respectively describe virtual crystal (conventional alloy) shifts to the CB, HH/LH, and SO band edge energies, as well as (vi) -- (vii) $a_{\scalebox{0.7}{\textrm{Bi}}}$ and $b_{\scalebox{0.7}{\textrm{Bi}}}$, the hydrostatic and axial deformation potentials associated with the Bi-related localised states. \cite{Broderick_SST_2013,Broderick_SST_2015} Using atomistic supercell electronic structure calculations we have explicitly computed the values of each of these parameters in GaAs$_{1-y}$Bi$_{y}$ and InAs$_{1-y}$Bi$_{y}$. \cite{Broderick_SST_2013} This provides a predictive model of the band structure evolution with Bi composition $y$ while circumventing the usual need to rely on post hoc fits to alloy experimental data (which generally allow fitting only to $\beta$ in the case of an estimated $\Delta E_{\scalebox{0.7}{\textrm{Bi}}}$, thereby introducing ambiguity into the parameter set).

To apply the 12-band Hamiltonian to quaternary In$_{x}$Ga$_{1-x}$As$_{1-y}$Bi$_{y}$ alloys we interpolate linearly between the Bi-related parameters computed for (In,Ga)As$_{1-y}$Bi$_{y}$. For the remaining parameters of the model, which are related to the conventional 8-band terms describing the In$_{x}$Ga$_{1-x}$As host matrix semiconductor, we use the values recommended by Vurgaftman and Meyer. \cite{Vurgaftman_JAP_2001} Details of this interpolation procedure, as well the full details of the parameters used in our calculations, can be found in Ref.~\onlinecite{Chai_SST_2015}. Our calculation of the band offsets in strained QW structures follows from an analytical diagonalisation of the bulk zone-centre 12-band Hamiltonian, details of which are provided in Ref.~\onlinecite{Broderick_SST_2015}.

The electronic structure of In$_{x}$Ga$_{1-x}$As$_{1-y}$Bi$_{y}$ QWs is calculated using the 12-band \textbf{k}$\cdot$\textbf{p} Hamiltonian in the envelope function and axial approximations. \cite{Meney_PRB_1994} Our numerical implementation is based on a reciprocal space plane wave expansion method, which provides a numerically robust and efficient approach to analyse the electronic and optical properties. \cite{Healy_JQE_2006,Ehrhardt_book_2014} Following the calculation of the QW band structure, our calculation of the optical gain proceeds at fixed temperature and injected carrier density by (i) directly utilising the calculated QW eigenstates to evaluate the transition matrix elements for inter-band optical transitions, \cite{Szmulowicz_PRB_1995} thereby explicitly accounting for the Bi-induced modifications to the electronic properties, (ii) using the optical transition matrix elements obtained in this manner to compute the spontaneous emission (SE) spectrum under the assumption of quasi-thermal carrier equilibrium, and (iii) transforming the SE spectrum to obtain the corresponding material gain spectrum. \cite{Chang_JSTQE_1995,Broderick_JSTQE_2015,Marko_SR_2016} Since we are concerned here solely with unstrained or compressively strained QWs which provide appreciable optical gain only for transverse electric- (TE-) polarised photons, we restict the SE calculation in (ii) to TE-polarised optical transitions in order to obtain the TE-polarised material gain following (iii). Based on our previous analysis of GaAs$_{1-x}$Bi$_{x}$/(Al)GaAs QW lasers, \cite{Marko_SR_2016} as well as the available spectroscopic data for In$_{x}$Ga$_{1-x}$As$_{1-y}$Bi$_{y}$ alloys, \cite{Petropoulos_APL_2011,Kudraweic_APL_2011,Zhong_APL_2012,Marko_APL_2012,Chai_SST_2015} we describe the Bi-induced inhomogenous spectral broadening of the optical spectra in our calculations using a hyperbolic secant lineshape of width 25 meV. \cite{Broderick_JSTQE_2015,Marko_SR_2016} All calculations are performed at $T = 300$ K.


\subsection{Laser modelling}
\label{sec:laser_modelling}

We focus our analysis on separate confinement heterostructure (SCH) devices, having In$_{x}$Ga$_{1-x}$As$_{1-y}$Bi$_{y}$ QWs surrounded by unstrained In$_{0.53}$Ga$_{0.47}$As barriers and InP cladding layers. This generic laser structure is illustrated schematically in Fig.~\ref{fig:structure_and_model}(a). To identify optimised laser structures we analyse the threshold characteristics, seeking structures which simultaneously minimise the threshold carrier density $n_{\scalebox{0.7}{\textrm{th}}}$ -- which can be expected to minimise the threshold current density -- and maximise the differential gain at threshold $\frac{dg}{dn}$. For the calculation of the threshold gain $g_{\scalebox{0.7}{\textrm{th}}}$ we assume a total cavity length of 1 mm and internal (cavity) losses of 5 cm$^{-1}$, chosen respectively to facilitate comparison to existing results on GaAs-based dilute bismide devices, \cite{Ludewig_APL_2013,Marko_SR_2016} and to reflect the low optical losses that should be achievable in a high quality InP-based laser structure. For each structure considered we optimise the optical confinement factor $\Gamma$ of the fundamental TE-polarised optical mode -- calculated using an effective index approach \cite{Kawano_book_2001,Chuang_book_2009} -- by varying the thickness of the barrier layers. Given the relatively low values of $\Gamma$ in mid-infrared SCH devices, we emphasise that choosing the barrier thickness in this manner is an important consideration for the design of these structures.

The choice of barrier and cladding layers considered here is motivated by the established epitaxial growth of In$_{x}$Ga$_{1-x}$As$_{1-y}$Bi$_{y}$ alloys: our intention is to identify structures to facilitate initial demonstration of an electrically pumped mid-infrared dilute bismide laser, with minimal disruption to established growth and fabrication processes. For example, it is possible to grow strain-compensated or strain-balanced structures incorporating tensile strained ternary or quaternary (Al)In$_{x}$Ga$_{1-x}$As(P) barriers. However, given (i) the large type-I band offsets achievable using unstrained In$_{0.53}$Ga$_{0.47}$As barriers, as well as (ii) that the optimised In$_{x}$Ga$_{1-x}$As$_{1-y}$Bi$_{y}$ QWs identified above lie safely within estimated strain-thickness limits, our analysis suggests limited benefit to pursuing such structures. In practice, inclusion of quaternary barrier layers complicates heterostructure growth -- e.g.~by requiring precise control of flux ratios for group-III and -V elements -- and could degrade material quality, while offering little tangible benefit in terms of carrier confinement and likely contributing to a degradation in optical confinement. Similarly, while employing more conventional lattice-matched Al$_{y}$In$_{x}$Ga$_{1-x-y}$As cladding layers may lead to improved threshold characteristics via enhancement of $\Gamma$, we note that this will have no impact on the calculated trends in the QW electronic and optical properties described here (aside from any reduction in $g_{\scalebox{0.7}{\textrm{th}}}$, and associated decrease (increase) in $n_{\scalebox{0.7}{\textrm{th}}}$ ($\frac{dg}{dn}$), brought about by any increase in $\Gamma$). Overall, we note that the conclusions reached on the basis of our optimisation of the active region of these simple structures are expected to hold in general for structures containing In$_{x}$Ga$_{1-x}$As$_{1-y}$Bi$_{y}$ QWs.


\section{Results}
\label{sec:results}

In this section we present the results of our theoretical investigation. We begin in Sec.~\ref{sec:InGaAsBi_band_structure} by firstly verifying the validity of our model for the In$_{x}$Ga$_{1-x}$As$_{1-y}$Bi$_{y}$ bulk band structure, and then apply this model to quantify the potential to engineer the band structure for applications in mid-infrared semiconductor lasers. In Sec.~\ref{sec:3500_nm_structures} we turn our attention to single QW laser structures designed to emit at 3.5 $\mu$m, eludicate general trends in the predicted properties of these devices, and identify structures which should deliver optimal performance. Then, in Sec.~\ref{sec:longer_wavelength} we discuss the dependence of the expected performance on the number of QWs and the emission wavelength, for structures designed to emit between 3 and 5 $\mu$m.


\subsection{\texorpdfstring{In$_{x}$Ga$_{1-x}$As$_{1-y}$Bi$_{y}$/InP}{InGaAsBi/InP} bulk band structure}
\label{sec:InGaAsBi_band_structure}

\begin{figure*}[t!]
	\includegraphics[width=0.95\textwidth]{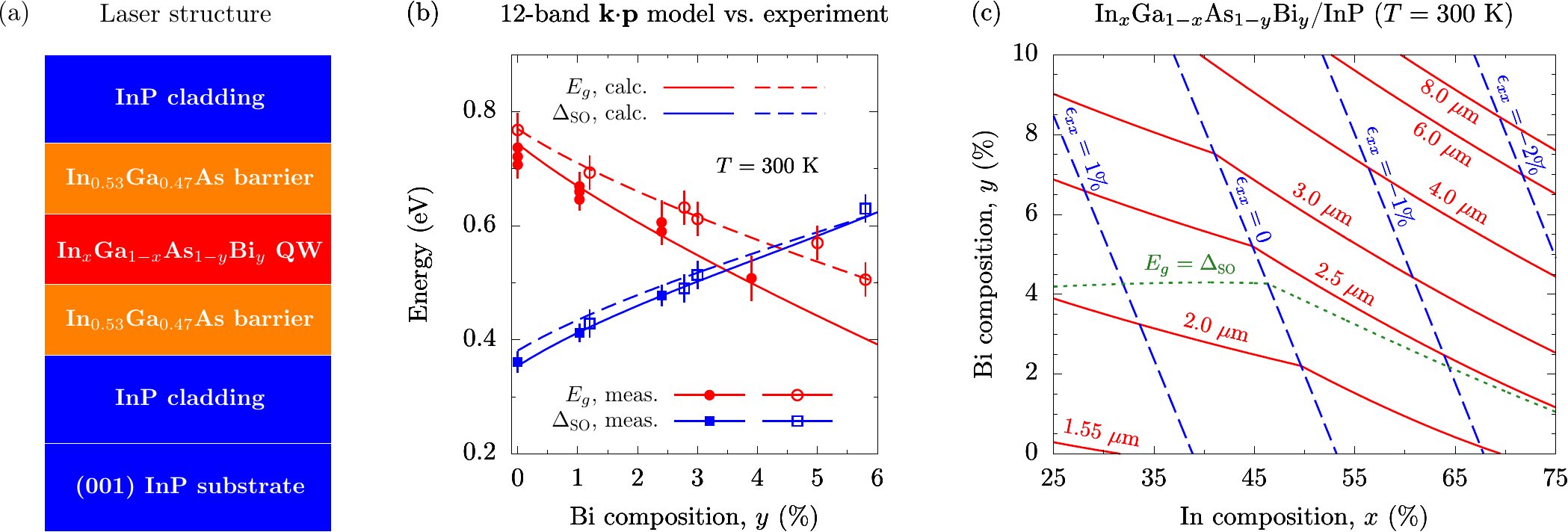}
	\caption{(a) Schematic illustration of the InP-based laser structures considered in this work: the structures consist of compressively strained In$_{x}$Ga$_{1-x}$As$_{1-y}$Bi$_{y}$ QWs surrounded by unstrained In$_{0.53}$Ga$_{0.47}$As barriers, with optical confinement provided by a SCH formed using InP cladding layers. (b) Measured variation of the band gap ($E_{g}$; closed and open red circles) and VB spin-orbit splitting energy ($\Delta_{\protect\scalebox{0.7}{\textrm{SO}}}$; closed and open blue squares) at $T = 300$ K, compared to that calculated using the 12-band \textbf{k}$\cdot$\textbf{p} model (solid and dashed red and blue lines, respectively) for pseudomorphically strained In$_{x}$Ga$_{1-x}$As$_{1-y}$Bi$_{y}$ bulk-like epitaxial layers grown on (001) InP. Results are shown for two sets of samples: the first are compressively strained In$_{0.53}$Ga$_{0.47}$As$_{1-y}$Bi$_{y}$ layers (closed circles/squares and solid lines), the second are In$_{x}$Ga$_{1-x}$As$_{1-y}$Bi$_{y}$ layers having fixed in-plane strain $\epsilon_{xx}$ (open circles/squares and dashed lines). (c) Composition space map illustrating the ranges of $\epsilon_{xx}$ and $E_{g}$ (equivalently, emission wavelength $\lambda$) accessible using pseudomorphically strained In$_{x}$Ga$_{1-x}$As$_{1-y}$Bi$_{y}$ bulk-like epitaxial layers grown on (001) InP. Dashed blue and solid red lines respectively denote paths in the composition space along which $\epsilon_{xx}$ and $\lambda$ are constant. The dotted green line denotes alloy compositions for which $E_{g} = \Delta_{\protect\scalebox{0.7}{\textrm{SO}}}$; alloys lying above this line have $\Delta_{\protect\scalebox{0.7}{\textrm{SO}}} > E_{g}$.}
	\label{fig:structure_and_model}
\end{figure*}

To demonstrate the validity of our parametrisation of the 12-band \textbf{k}$\cdot$\textbf{p} Hamiltonian for quaternary In$_{x}$Ga$_{1-x}$As$_{1-y}$Bi$_{y}$ alloys we have calculated the evolution of $E_{g}$ and $\Delta_{\scalebox{0.7}{\textrm{SO}}}$ with Bi composition $y$. Figure~\ref{fig:structure_and_model}(b) compares the results of these calculations to a range of experimental measurments -- undertaken at room temperature using photoluminescence, photo-modulated reflectance and absorption spectroscopy -- for two distinct sets of In$_{x}$Ga$_{1-x}$As$_{1-y}$Bi$_{y}$ bulk-like epitaxial layers grown pseudmorphically on InP. Full details of the growth and characterisation of these samples, as well as details of the experimental measurements, can be found in Refs.~\onlinecite{Chai_SST_2015}, \onlinecite{Petropoulos_APL_2011} and~\onlinecite{Zhong_APL_2012}.

The first set of samples are compressively strained In$_{0.53}$Ga$_{0.47}$As$_{1-y}$Bi$_{y}$ epitaxial layers; the measured and calculated values of $E_{g}$ are shown respectively using closed red circles and solid red lines, while the measured and calculated values of $\Delta_{\scalebox{0.7}{\textrm{SO}}}$ are shown respectively using closed blue squares and solid blue lines. \cite{Kudraweic_APL_2011,Chai_SST_2015} We find that the predicted decrease (increase) and strong composition-dependent bowing of $E_{g}$ ($\Delta_{\scalebox{0.7}{\textrm{SO}}}$) with $y$ predicted by the 12-band model is in good quantitative agreement with experiment. For this set of samples we calculate that $E_{g}$ decreases by 74 meV, and $\Delta_{\scalebox{0.7}{\textrm{SO}}}$ increases by 58 meV, when 1\% of the As atoms in In$_{0.53}$Ga$_{0.47}$As are replaced by Bi. On this basis, our calculations predict that a bulk band structure in which $\Delta_{\scalebox{0.7}{\textrm{SO}}} > E_{g}$ can be achieved in compressively strained In$_{x}$Ga$_{1-x}$As$_{1-y}$Bi$_{y}$/InP pseudomorphic layers for Bi compositions as low as 3.5\%. \cite{Broderick_PSSB_2013,Jin_JAP_2013}


The second set of samples are In$_{x}$Ga$_{1-x}$As$_{1-y}$Bi$_{y}$ epitaxial layers having fixed in-plane strain $\epsilon_{xx}$ with respect to the InP substrate upon which they were grown. \cite{Zhong_APL_2012} The measured and calculated values of $E_{g}$ for this set of samples are shown respectively using open red circles and dashed red lines, while the corresponding values of $\Delta_{\scalebox{0.7}{\textrm{SO}}}$ are shown respectively using open blue squares and dashed blue lines. Nominally these samples are lattice-matched to InP, having $\epsilon_{xx} = 0$. However, examination of the experimental data for the Bi-free ($y = 0$) sample shows that the measured value of $E_{g}$ is higher than that which one would expect for lattice-matched In$_{0.53}$Ga$_{0.47}$As, suggesting the presence of a slightly lower In composition. To analyse these samples we therefore begin by determining the In composition required to produce the measured room temperature band gap at $y = 0$, and obtain an In composition $x = 43$\%. This In composition, which is 10\% lower than the 53\% required for lattice-matching to InP, corresponds to an in-plane tensile strain $\epsilon_{xx} = 0.7$\%. This is consistent with high resolution x-ray diffraction measurements carried out on these samples, which indicate the presence of small tensile strains rather than exact lattice matching. \cite{Zhong_APL_2012} Given that this set of samples has fixed strain with respect to the InP substrate, we then calculate the variation of $E_{g}$ and $\Delta_{\scalebox{0.7}{\textrm{SO}}}$ with $y$ by simultaneously reducing $x$ so that $\epsilon_{xx} = 0.7$\% is maintained fixed. Again we find good quantitative agreement with the experimental data, with the calculated reduction in $E_{g}$ per \% Bi replacing As slightly reduced compared to that in In$_{0.53}$Ga$_{0.47}$As$_{1-y}$Bi$_{y}$ (a consequence of the reduction in $x$ required to maintain $\epsilon_{xx}$ fixed with increasing $y$). \cite{Broderick_bismide_chapter_2017}

We note that the measured data for the set of samples having fixed strain indicate that slightly higher Bi compositions $y \approx 4.5$\% are required to bring about $\Delta_{\scalebox{0.7}{\textrm{SO}}} > E_{g}$. \cite{Chai_SST_2015,Broderick_bismide_chapter_2017} The available experimental data for the In$_{x}$Ga$_{1-x}$As$_{1-y}$Bi$_{y}$ sample having $y = 5.8$\% -- the highest Bi composition for which experimental data are currently available -- clearly demonstrate that a band structure having $\Delta_{\scalebox{0.7}{\textrm{SO}}} > E_{g}$ can be readily achieved in this material system. \cite{Chai_SST_2015} Further, temperature-dependent spectroscopic measurements on this sample have confirmed that $\Delta_{\scalebox{0.7}{\textrm{SO}}} > E_{g}$ is maintained at all temperatures. Overall, we conclude that our parametrisation of the 12-band model for quaternary In$_{x}$Ga$_{1-x}$As$_{1-y}$Bi$_{y}$ alloys is in good agreement with the available experimental data and, as such, should provide a reliable basis for investigation of the properties and performance of In$_{x}$Ga$_{1-x}$As$_{1-y}$Bi$_{y}$ laser structures.


Having demonstrated the validity of our theoretical model of the band structure, we turn our attention now to the potential to engineer the In$_{x}$Ga$_{1-x}$As$_{1-y}$Bi$_{y}$ band structure for applications in mid-infrared light-emitting devices. The composition space map in Fig.~\ref{fig:structure_and_model}(c) depicts the ranges of in-plane strain $\epsilon_{xx}$ and room temperature band gap $E_{g}$ (emission wavelength $\lambda$) accessible using In$_{x}$Ga$_{1-x}$As$_{1-y}$Bi$_{y}$ bulk-like epitaxial layers grown pseudmorphically on InP. \cite{Broderick_NUSOD_2016_InP,Broderick_bismide_chapter_2017} Dashed blue and solid red lines in Fig.~\ref{fig:structure_and_model}(c) respectively denote paths in the compostion space along which $\epsilon_{xx}$ and $\lambda$ are constant. The dotted green line denotes alloy compositions for which $\Delta_{\scalebox{0.7}{\textrm{SO}}} = E_{g}$, with alloys lying above this line having $\Delta_{\scalebox{0.7}{\textrm{SO}}} > E_{g}$. These calculations suggest that a broad range of emission wavelengths can be accessed using In$_{x}$Ga$_{1-x}$As$_{1-y}$Bi$_{y}$ alloys, ranging from $\approx 1.55$ to $> 3$ $\mu$m in tensile strained ($\epsilon_{xx} > 0$) layers, and from $\approx 2$ to $\gtrsim 8$ $\mu$m in compressively strained ($\epsilon_{xx} < 0$) layers. The wavelength range accessible using tensile strained In$_{x}$Ga$_{1-x}$As$_{1-y}$Bi$_{y}$ QWs is significantly curtailed compared to that accessible using compressively strained QWs. For example, for a fixed tensile strain $\epsilon_{xx} = 1.5$\% we find, when quantum confinement effects are taken into account, that it is not possible to achieve $\lambda \gtrsim 3$ $\mu$m.

Using Voisin's expression for the critical thickness $t_{c}$ of a pseudmorphically strained layer, \cite{Voisin_SPIE_1988,Reilly_SST_1989} we calculate a strain-thickness limit $t_{c} \vert \epsilon_{xx} \vert \approx 22.3$ nm \% for a compressively strained In$_{x}$Ga$_{1-x}$As epitaxial layer grown on (001)-oriented InP. Incorporation of Bi slightly softens the material by reducing the elastic constants $C_{11}$ and $C_{12}$, leading to a slight increase in $t_{c}$ for an (001)-oriented In$_{x}$Ga$_{1-x}$As$_{1-y}$Bi$_{y}$ epitaxial layer. However, since we are concerned here with dilute Bi compositions $y \lesssim 10$\%, the associated increase in the strain-thickness limit is minimal. As such, we expect that the calculated strain-thickness limit for compressively strained In$_{x}$Ga$_{1-x}$As should serve as a reliable estimate for the compressively strained In$_{x}$Ga$_{1-x}$As$_{1-y}$Bi$_{y}$ QWs considered here. On this basis, we estimate that 10 nm thick In$_{x}$Ga$_{1-x}$As layers can be grown on InP up to a compressive strain $\vert \epsilon_{xx} \vert \approx 2.2$\% before the onset of plastic relaxation.

The ability to independently vary $\epsilon_{xx}$ and $\lambda$ in quaternary In$_{x}$Ga$_{1-x}$As$_{1-y}$Bi$_{y}$ alloys presents significant scope to engineer the band structure, in order to obtain properties suitable for the development of high performance laser devices. In particular, our analysis suggests that (i) emission between 3 and 5 $\mu$m can be acheived for In (Bi) compositions ranging from approximately $x = 55$ -- 70\% ($y = 4$ -- 8\%), (ii) that this corresponds to compressive strains ranging from $\epsilon_{xx} = 0$ -- 2\% in alloys having $\Delta_{\scalebox{0.7}{\textrm{SO}}} > E_{g}$, which can be expected to have significant positive impact on the laser performance, \cite{Sweeney_IEEEISLC_2010,Sweeney_ICTON_2011,Broderick_SST_2012} (iii) that QWs based on these alloys have large type-I band offsets (cf.~Sec.~\ref{sec:3500_nm_structures}), which can be expected to minimise losses associated with thermal carrier leakage at high temperatures, and (iv) that suitable QWs for such applications can be grown within estimated strain-thickness limits.

We note that for $\lambda$ ranging from approximately 2 -- 3 $\mu$m, compressively strained In$_{x}$Ga$_{1-x}$As$_{1-y}$Bi$_{y}$ alloys have $\Delta_{\scalebox{0.7}{\textrm{SO}}} \approx E_{g}$. Such a band structure can be expected to significantly limit device performance due to IVBA and CHSH Auger recombination processes involving the SO VB being close to resonant. Similarly, we expect the performance of tensile strained In$_{x}$Ga$_{1-x}$As$_{1-y}$Bi$_{y}$ devices having $\lambda \lesssim 2$ $\mu$m to be strongly limited by IVBA and CHSH Auger recombination, due to their possessing a band structure in which $\Delta_{\scalebox{0.7}{\textrm{SO}}} \lesssim E_{g}$ with $E_{g}$ and $\Delta_{\scalebox{0.7}{\textrm{SO}}}$ being relatively close in magnitude. Overall, our calculations then suggest that In$_{x}$Ga$_{1-x}$As$_{1-y}$Bi$_{y}$ alloys are best suited for mid-infrared applications at $\lambda \gtrsim 3$ $\mu$m, with the wavelength range accessible above 3 $\mu$m likely to be limited by a combination of material growth (i.e.~what combinations of $x$ and $y$ can be achieved in materials having sufficiently high quality to facilitate electrically-pumped lasing), and hot-electron producing CHCC Auger recombination (the rate of which increases strongly with increasing $\lambda$, \cite{Silver_JQE_1997} and which we expect to be the dominant loss mechanism in these devices).


Finally, we emphasise that the Bi compositions at which $\Delta_{\scalebox{0.7}{\textrm{SO}}} > E_{g}$ can be achieved are significantly lower in In$_{x}$Ga$_{1-x}$As$_{1-y}$Bi$_{y}$ than the $\approx 10$\% Bi required to achieve the same crossover in GaAs$_{1-y}$Bi$_{y}$/GaAs alloys. \cite{Usman_PRB_2011,Broderick_SST_2013,Broderick_JSTQE_2015} As such, our analysis indicates that it should be feasible to achieve $\Delta_{\scalebox{0.7}{\textrm{SO}}} > E_{g}$ in mid-infrared In$_{x}$Ga$_{1-x}$As$_{1-y}$Bi$_{y}$ laser structures. Indeed, all structures considered in our analysis in Secs.~\ref{sec:3500_nm_structures} and~\ref{sec:longer_wavelength} satisfy this condition. While a band structure having $\Delta_{\scalebox{0.7}{\textrm{SO}}} > E_{g}$ is readily achievable in GaSb-based devices emitting beyond 2 $\mu$m, \cite{Sweeney_IEEEISLC_2010,Sweeney_ICTON_2011,Eales_JSTQE_2017} utilising In$_{x}$Ga$_{1-x}$As$_{1-y}$Bi$_{y}$ QWs enables this to be achieved in InP-based devices. Our analysis therefore suggests that it should be possible to mitigate losses related to IVBA and CHSH Auger recombination, and that this can in principle be achieved more readily than in GaAs-based dilute bismide laser structures due to the significantly reduced Bi compositions required. As such, In$_{x}$Ga$_{1-x}$As$_{1-y}$Bi$_{y}$ structures should provide a useful basis upon which to demonstrate the reduction of such losses in dilute bismide-based heterostructures. Given the potential to exploit this band structure condition to realise semiconductor lasers having high efficiency and temperature stability, this has the potential to act as an important proof of principle for GaAs-based near-infrared devices, where efforts are ongoing to achieve emission out to $\lambda = 1.55$ $\mu$m ($y \approx 13$\%) in structures having $\Delta_{\scalebox{0.7}{\textrm{SO}}} > E_{g}$ in the active (QW) region.


\subsection{Analysis and optimisation of single quantum well laser structures at \texorpdfstring{$\lambda = 3.5$ $\mu$m}{3.5 micron}}
\label{sec:3500_nm_structures}

The use of a quaternary alloy to form the QW layer(s) of the laser structure depicted in Fig.~\ref{fig:structure_and_model}(a) provides significant freedom to engineer the band structure, allowing $\epsilon_{xx}$ and $\lambda$ to be varied independently. For a desired $\lambda$, this then allows for the growth of an extremely large number of structures having distinct combinations of alloy composition $x$ and $y$, in-plane strain $\epsilon_{xx}$, and QW thickness $t$. However, the performance of a given laser structure is in general largely determined by the interplay of a small number of complementary and competing factors: (i) the density of states (DOS) close to the VB edge, (ii) the inter-band optical transition strengths, and (iii) the optical confinement factor $\Gamma$. \cite{Reilly_JQE_1994,Adams_JSTQE_2011} In practice, each of these factors is determined in large part by a combination of distinct QW properties, allowing for general trends to be readily quantified. Firstly, the DOS close to the VB edge is largely determined by $\epsilon_{xx}$ and $t$, since compressive (tensile) strain acts to decrease (increase) the in-plane hole effective masses, and the larger confinement energy in narrower QWs acts to increase the separation in energy between subbands. Secondly, the inter-band optical transition strengths are primarily governed by a combination of the band offsets and $t$, which determine the nature of the carrier localisation and hence the spatial overlap between bound electron and hole states. Thirdly, $\Gamma$ is determined by a combination of $t$, the refractive index constrast between the barrier and cladding layers, and the thickness of the barrier and cladding layers.

In order to analyse the properties of In$_{x}$Ga$_{1-x}$As$_{1-y}$Bi$_{y}$ QWs for a given emission wavelength $\lambda$, we focus our attention on two distinct sets of structures. Firstly, we consider QWs having fixed thickness and variable compressive strain: we choose $t = 7$ nm and vary $\epsilon_{xx}$ from 0 to $-2$\% by adjusting $x$ and $y$ in the QW to maintain fixed $\lambda$. Secondly, we consider QWs having fixed compressive strain and variable thickness: we choose $\epsilon_{xx} = -1.5$\% and vary $t$ from 3 to 10 nm, again adjusting $x$ and $y$ to maintain fixed $\lambda$. In addition to quantifying key trends in the expected performance, this approach enables us to identify specific laser structures offering optimal performance -- i.e.~QWs which simultaneously minimise $n_{\scalebox{0.7}{\textrm{th}}}$ and maximise $\frac{dg}{dn}$. \cite{Silver_IEEEJQE_1995} We have performed this systematic analysis for structures designed to emit between 3 and 5 $\mu$m, and find that the same general qualitative trends determine the calculated properties throughout this wavelength range. As such, we present here a full description of the results of these calculations only for structures having $\lambda = 3.5$ $\mu$m. We then describe the calculated dependence of the laser performance on $\lambda$, as well as the number of QWs in multi-QW structures, in Sec.~\ref{sec:longer_wavelength}.

We begin our optimisation by varying the thickness of the In$_{0.53}$Ga$_{0.47}$As barrier layers in order to maximise $\Gamma$. For the laser structure depicted in Fig.~\ref{fig:structure_and_model}(a), we calculate that this is achieved at $\lambda = 3.5$ $\mu$m for structures having 450 nm thick barriers both below and above the QW, resulting in $\Gamma = 0.53$\% for a single QW structure having $t = 7$ nm. As such, we maintain this fixed barrier thickness for our calculation of $\Gamma$ and the threshold material gain $g_{\scalebox{0.7}{\textrm{th}}}$ for all structures desiged to emit at 3.5 $\mu$m.


\subsubsection{Fixed thickness, variable strain}
\label{sec:variable_strain}

\begin{figure*}[t!]
	\includegraphics[width=1.00\textwidth]{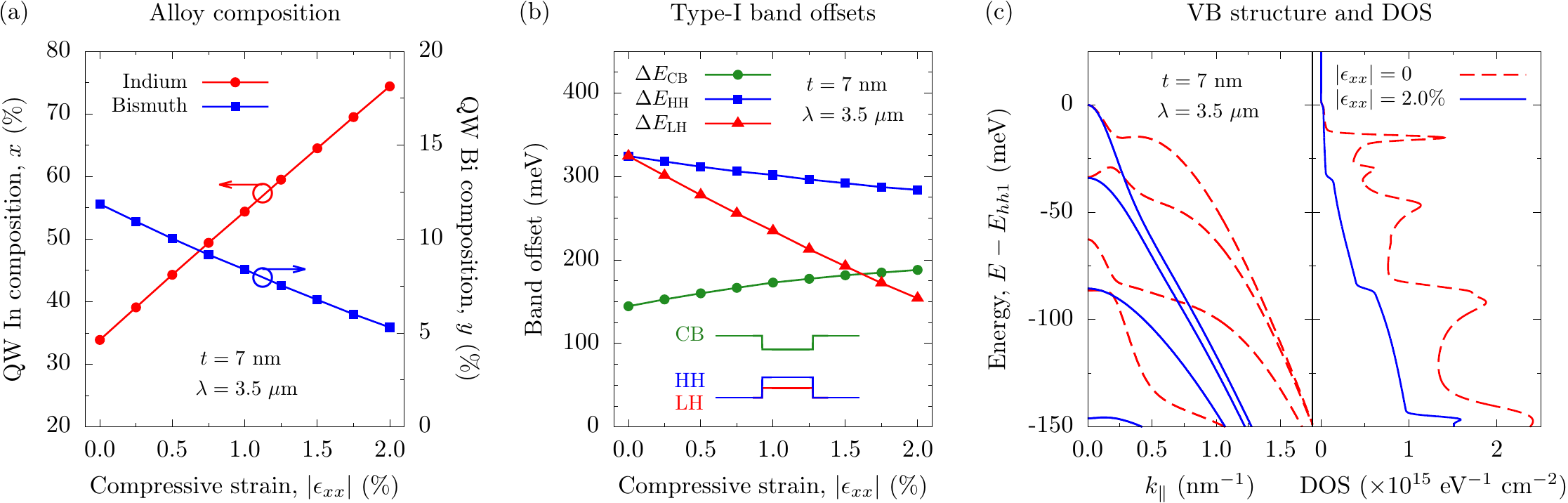}
	\caption{Summary of calculated electronic properties for In$_{x}$Ga$_{1-x}$As$_{1-y}$Bi$_{y}$ QWs having fixed thickness $t = 7$ nm, variable compressive in-plane strain $\epsilon_{xx}$, and emission wavelength $ \lambda = 3.5$ $\mu$m at $T = 300$ K. (a) In$_{x}$Ga$_{1-x}$As$_{1-y}$Bi$_{y}$ QW alloy In and Bi compositions $x$ and $y$ required to maintain $\lambda = 3.5$ $\mu$m as a function of the magnitude $\vert \epsilon_{xx} \vert$ of the compressive strain. (b) Type-I CB (closed green circles), HH (closed blue circles) and LH (closed red triangles) band offsets, between the In$_{x}$Ga$_{1-x}$As$_{1-y}$Bi$_{y}$ QW and In$_{0.53}$Ga$_{0.47}$As barrier layers, as a function of $\vert \epsilon_{xx} \vert$. (c) VB structure (left panel) and DOS (right panel) for the QWs having $\epsilon_{xx} = 0$ (unstrained; dashed red lines) and $\epsilon_{xx} = -2$\% (compressively strained; solid blue lines).}
     \label{fig:fixed_thickness_electronic}
\end{figure*}
\begin{figure*}[t!]
	\includegraphics[width=1.00\textwidth]{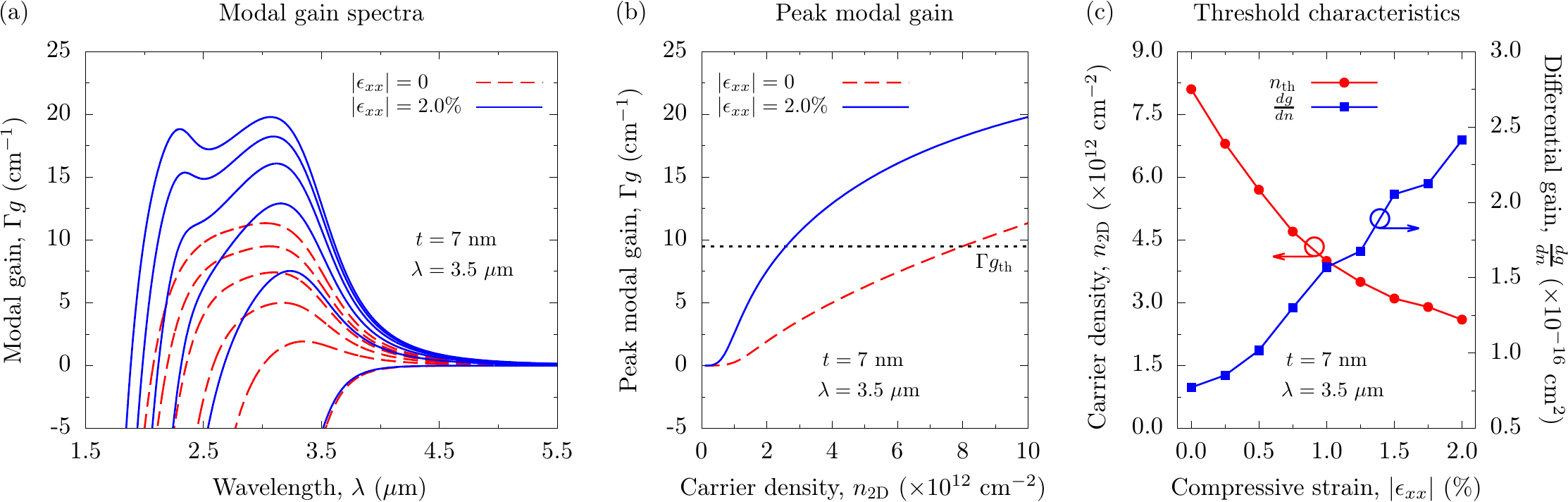}
	\caption{Summary of calculated optical properties for In$_{x}$Ga$_{1-x}$As$_{1-y}$Bi$_{y}$ QWs having fixed thickness $t = 7$ nm, variable compressive in-plane strain $\epsilon_{xx}$, and emission wavelength $ \lambda = 3.5$ $\mu$m at $T = 300$ K. (a) TE-polarised modal gain spectra for the QWs having $\epsilon_{xx} = 0$ (dashed red lines) and $\epsilon_{xx} = -2$\% (solid blue lines), calculated at sheet carrier densities $n_{\protect\scalebox{0.7}{\textrm{2D}}} = 10^{11}$ -- $10^{13}$ cm$^{-2}$ (in increments of $2 \times 10^{12}$ cm$^{-2}$). (b) Peak TE-polarised modal gain $\Gamma g$ as a function of $n_{\protect\scalebox{0.7}{\textrm{2D}}}$ for the same two QWs as in (a). The horizontal dotted black line denotes the calculated threshold modal gain, $\Gamma g_{\protect\scalebox{0.7}{\textrm{th}}} = 9.5$ cm$^{-1}$. (c) Threshold sheet carrier density ($n_{\protect\scalebox{0.7}{\textrm{th}}}$; closed red circles) and differential material gain at threshold ($\protect\frac{dg}{dn}$; closed blue squares) as a function of $\vert \epsilon_{xx} \vert$.}
     \label{fig:fixed_thickness_optical}
\end{figure*}

The results of our calculations for $\lambda = 3.5$ $\mu$m laser structures having fixed $t$ and variable $\epsilon_{xx}$ are summarised in Figs.~\ref{fig:fixed_thickness_electronic} and~\ref{fig:fixed_thickness_optical}. Figure~\ref{fig:fixed_thickness_electronic}(a) shows the required variation in $x$ (closed red circles) and $y$ (closed blue squares) required to achieve $\lambda = 3.5$ $\mu$m -- corresponding to an $e1$-$hh1$ transition energy 0.354 eV -- as $\epsilon_{xx}$ is varied from 0 to $-2$\% in a QW of fixed thickness $t = 7$ nm. For the unstrained structure we calculate that respective In and Bi compositions $x = 33.9$\% and $y = 11.9$\% are required to maintain $\lambda = 3.5$ $\mu$m. As the magnitude $\vert \epsilon_{xx} \vert$ of the compressive strain increases from 0 to 2\%, the In (Bi) composition required to maintain $\lambda = 3.5$ $\mu$m increases (decreases) approximately linearly, to $x = 74.4$\% ($y = 5.3$\%). As a result of the fixed thickness of this set of QWs, there is little variation in the total confinement energy of the lowest energy bound electron state $e1$ and highest energy hole state $hh1$ (which is HH-like at $k_{\parallel} = 0$ in all QWs investigated). We calculate that this confinement energy remains approximately constant at 83 meV between $\vert \epsilon_{xx} \vert = 0$ and 2\%, so that this set of structures requires In$_{x}$Ga$_{1-x}$As$_{1-y}$Bi$_{y}$ QW layers having a fixed bulk room temperature band gap of approximately 0.271 eV (4.58 $\mu$m). Recalling Fig.~\ref{fig:structure_and_model}(c), these QWs then have alloy compositions lying on a contour running approximately parallel to the 4.0 $\mu$m contour (solid red line), but at slightly higher $x$ and $y$.


The calculated variation of the type-I CB, HH and LH band offsets $\Delta E_{\scalebox{0.7}{\textrm{CB}}}$, $\Delta E_{\scalebox{0.7}{\textrm{HH}}}$ and $\Delta E_{\scalebox{0.7}{\textrm{LH}}}$ with $\vert \epsilon_{xx} \vert$ -- between the bulk band edge energies of the In$_{x}$Ga$_{1-x}$As$_{1-y}$Bi$_{y}$ QW and In$_{0.53}$Ga$_{0.47}$As barrier layers -- for these QWs are shown in Fig.~\ref{fig:fixed_thickness_electronic}(b) using, respectively, closed green circles, blue squares and red triangles. In the unstrained QW the HH and LH bulk VB edge states of the In$_{0.339}$Ga$_{0.661}$As$_{0.881}$Bi$_{0.119}$ QW layer are degenerate and, since these QWs have unstrained barriers, the HH and LH band offsets are equal. We calculate $\Delta E_{\scalebox{0.7}{\textrm{HH}}} = \Delta E_{\scalebox{0.7}{\textrm{LH}}} = 325$ meV in the unstrained QW.


As $\vert \epsilon_{xx} \vert$ is increased from 0 to 2\% we calculate (i) that the degeneracy of the HH and LH bulk VB edge states in the In$_{x}$Ga$_{1-x}$As$_{1-y}$Bi$_{y}$ QW layer is lifted in the usual manner, by the axial component of the pseudomorphic strain, and (ii) that both $\Delta E_{\scalebox{0.7}{\textrm{HH}}}$ and $\Delta E_{\scalebox{0.7}{\textrm{LH}}}$ decrease in magnitude. The calculated, roughly linear, decrease in both $\Delta E_{\scalebox{0.7}{\textrm{HH}}}$ and $\Delta E_{\scalebox{0.7}{\textrm{LH}}}$ with increasing $\vert \epsilon_{xx} \vert$ is primarily associated with the decrease in $y$, with the axial component of the strain acting to respectively mitigate and enhance this reduction for $\Delta E_{\scalebox{0.7}{\textrm{HH}}}$ and $\Delta E_{\scalebox{0.7}{\textrm{LH}}}$. For the In$_{0.744}$Ga$_{0.256}$As$_{0.947}$Bi$_{0.053}$ QW having $\epsilon_{xx} = -2$\% strain we calculate $\Delta E_{\scalebox{0.7}{\textrm{HH}}} = 284$ meV and $\Delta E_{\scalebox{0.7}{\textrm{LH}}} = 154$ meV. For the CB offset $\Delta E_{\scalebox{0.7}{\textrm{CB}}}$, we calculate a value of 145 meV in the unstrained In$_{0.339}$Ga$_{0.661}$As$_{0.881}$Bi$_{0.119}$ QW, which increases to 188 meV in the In$_{0.744}$Ga$_{0.256}$As$_{0.947}$Bi$_{0.053}$ QW having $\epsilon_{xx} = -2$\%. We find that the calculated increase in $\Delta E_{\scalebox{0.7}{\textrm{CB}}}$ is associated with the increase in $x$, with the downward shift of the bulk CB edge energy in the QW layer sufficient to overcome the upward shifts due to (i) the hydrostatic component of the compressive strain, and (ii) the reduction in $y$.

To quantify the impact of the band offsets on the expected laser performance we have calculated (i) the fraction of the total electron and hole charge density confined within the QW at threshold (i.e.~at an injected carrier density $n_{\scalebox{0.7}{\textrm{th}}}$), \cite{Healy_JQE_2006} (ii) the ionisation energies for the escape of bound electron and hole states from the QW, and (iii) the inter-band optical matrix elements for transitions between the bound $e1$ and $hh1$ electron and hole states at the centre of the QW Brillouin zone ($k_{\parallel} = 0$). We describe transition strengths associated with the latter in units of energy, with the relevant scale being the Kane parameter $E_{P}$. \cite{Szmulowicz_PRB_1995,Broderick_JSTQE_2015,Broderick_bismide_chapter_2017}

At threshold, we calculate that 38.9\% (98.1\%) and 66.9\% (97.6\%) of the total electon (hole) charge density resides within the QW for $\epsilon_{xx} = 0$ and $-2$\%, respectively, with these QWs having $n_{\scalebox{0.7}{\textrm{th}}} = 8.1$ and $2.6 \times 10^{12}$ cm$^{-2}$ (cf.~Fig.~\ref{fig:fixed_thickness_optical}(c)). These results suggest excellent confinement and negligible thermal spill-out of holes at room temperature, but that significant thermal spill-out of electrons can be expected in unstrained QWs. These trends are confirmed by considering the variation with $\vert \epsilon_{xx} \vert$ of the ionisation energy of an $e1$ electron, calculated as the difference in energy between the bound $e1$ state at $k_{\parallel} = 0$ and the bulk CB edge in the surrounding In$_{0.53}$Ga$_{0.47}$As barrier layers. At $\epsilon_{xx} = 0$ we calculate an ionisation energy of 75 meV, which increases approximately linearly to 115 meV as $\vert \epsilon_{xx} \vert$ increases to 2\%, in line with the calculated increase in $\Delta E_{\scalebox{0.7}{\textrm{CB}}}$. For this series of QWs the ionisation energy for a zone-centre $e1$ electron is then at least three times larger than the thermal energy $k_{\scalebox{0.7}{\textrm{B}}} T = 26$ meV at room temperature, indicating that the calculated spill-out of the electron charge density at threshold in the unstrained QW is associated with the occupation of (i) a significant number of states in the $e1$ subbands away from $k_{\parallel} = 0$, and (ii) occupation of higher energy conduction subbands, both of which are brought about by the high value of $n_{\scalebox{0.7}{\textrm{th}}}$ for this structure. The corresponding $hh1$ hole ionisation energy decreases with increasing $\vert \epsilon_{xx} \vert$, in line with the calculated reduction in $\Delta E_{\scalebox{0.7}{\textrm{HH}}}$, but remains in excess of 270 meV at $\vert \epsilon_{xx} \vert = 2$\%. The calculated improvement in electron confinement in going from 0 to 2\% compressive strain indicates that it is particularly desirable to grow strained QWs to mitigate thermal leakage of electrons, which may otherwise limit performance at and above room temperature.

Considering now the optical matrix elements between the $e1$ and $hh1$ states, we calculate respective transition strengths of 12.42 and 13.39 eV at $\epsilon_{xx} = 0$ and $-2$\%. Our analysis indicates that this $\approx 8$\% increase in going from $\epsilon_{xx} = 0$ to $-2$\% is primarily associated with the improved confinement of the $e1$ envelope function, brought about by the calculated 43 meV increase in $\Delta E_{\scalebox{0.7}{\textrm{CB}}}$. These values of the optical transition strengths are approximately two-thirds of those we have calculated for metamorphic InAs$_{1-x}$Sb$_{x}$/Al$_{0.125}$In$_{0.875}$As QWs having $\lambda \approx 3.5$ $\mu$m. \cite{Menendez_submitted_2018} We recall that the reduced optical transition strengths in Bi-containing heterostructures are brought about by the QW VB states consisting of a VBAC-induced admixture of In$_{x}$Ga$_{1-x}$As-derived extended and Bi-related localised states, the latter of which do not couple optically to the In$_{x}$Ga$_{1-x}$As-derived CB states. \cite{Broderick_JSTQE_2015,Broderick_bismide_chapter_2017}

Further analysis highlights that the calculated thermal spill-out of the electron charge density in the unstrained QW arises due to the high value of $n_{\scalebox{0.7}{\textrm{th}}}$ for this structure. This high level of injection brings about partial occupation of higher energy electron subbands, whose envelope functions are significantly more delocalised than those associated with $e1$. This is mitigated as $n_{\scalebox{0.7}{\textrm{th}}}$ decreases strongly with increasing $\vert \epsilon_{xx} \vert$, with a large majority ($> 75$\% at $\vert \epsilon_{xx} \vert = 2.0$\%) of injected electrons occupying the $e1$ subbands at threshold. Our calculated values of $\Delta E_{\scalebox{0.7}{\textrm{CB}}}$ and $\Delta E_{\scalebox{0.7}{\textrm{HH}}}$ are in excess of 140 meV in all structures considered -- i.e.~at least five times larger than the typical carrier thermal energy $k_{\scalebox{0.7}{\textrm{B}}} T$ at room temperature. On the basis of our analysis of carrier localisation in these structures we predict that (i) these band offsets should be sufficient to largely mitigate thermal carrier leakage in QWs designed to minimise $n_{\scalebox{0.7}{\textrm{th}}}$, and (ii) any thermal leakage from these structures will be associated almost entirely with thermionic emission of electrons.


The left- and right-hand panels in Fig.~\ref{fig:fixed_thickness_electronic}(c) respectively show the calculated VB structure and DOS for the In$_{0.339}$Ga$_{0.661}$As$_{0.881}$Bi$_{0.119}$ ($\epsilon_{xx} = 0$; dashed red lines) and In$_{0.744}$Ga$_{0.256}$As$_{0.947}$Bi$_{0.053}$ ($\epsilon_{xx} = -2.0$\%; solid blue lines) QWs. In order to compare the band dispersion and DOS, the zero of energy in each case has been chosen at the energy of the highest energy hole state $hh1$ at $k_{\parallel} = 0$. Comparing the VB stucture in both cases, we note that the compressive strain in the In$_{0.744}$Ga$_{0.256}$As$_{0.947}$Bi$_{0.053}$ QW produces significantly reduced in-plane hole effective masses. Of the five highest energy bound hole states we calculate that four are HH-like at $k_{\parallel} = 0$ ($hh1$ -- $hh4$) between $\epsilon_{xx} = 0$ and $-2$\%, with one being primarily LH-like ($lh1$). At $\epsilon_{xx} = 0$ and $-2$\% we respectively calculate that the third and fifth highest energy hole state is primarily LH-like at $k_{\parallel} = 0$. As such, we expect there to be minimal transverse magnetic- (TM-) polarised optical gain in QWs having compressive strains $\vert \epsilon_{xx} \vert \gtrsim 1$\%, so that the majority of injected holes contribute to the TE-polarised optical mode.

Turning our attention to the VB DOS, we note that the reduction of the in-plane hole effective masses at $\epsilon_{xx} = -2$\% leads to a significant reduction of the DOS close in energy to the VB edge. This is of primary importance in optimising the performance of these devices. Indeed, utilising compressive strain to reduce the VB DOS ensures that holes occupy states over a smaller (larger) range of $k_{\parallel}$ (energy) than in an equivalent unstrained QW. This more closely matches the distribution of VB edge holes to that of CB edge electrons, making more holes available to contribute to the optical gain. Correspondingly, the reduction in the DOS at the VB edge pushes the hole quasi-Fermi level downwards in energy at fixed injection, thereby increasing the separation $\Delta F$ between the electron and hole quasi-Fermi levels at fixed $n_{\scalebox{0.7}{\textrm{2D}}}$, and leading to decreased $n_{\scalebox{0.7}{\textrm{th}}}$ and increased $\frac{dg}{dn}$. The calculated strong reduction in the VB edge DOS again indicates that the growth of compressively strained QWs is essential to optimise performance. \cite{Adams_EL_1986,Reilly_JQE_1994,Adams_JSTQE_2011}


Figure~\ref{fig:fixed_thickness_optical}(a) shows the TE-polarised modal gain spectra $\Gamma g$ for the same $\epsilon_{xx} = 0$ and $-2$\% QWs, calculated for sheet carrier densities $n_{\scalebox{0.7}{\textrm{2D}}} = 0.1$, 2, 4, 6, 8 and $10 \times 10^{12}$ cm$^{-2}$. Here, we can clearly see the impact of the reduction in the VB edge DOS brought about by compressive strain. Firstly, we consider the transparency ($g = 0$) points on the low wavelength (high energy) side of the gain peak which, at fixed injection, occur when the photon energy is equal to the quasi-Fermi level separation $\Delta F$ (i.e.~at wavelength $\lambda = \frac{hc}{\Delta F}$). Comparing the calculated variation of $\Delta F$ with $n_{\scalebox{0.7}{\textrm{2D}}}$ at transparency for the unstrained (dashed red lines) and $\epsilon_{xx} = -2$\% (solid blue lines) QWs, we note that the reduction in the VB edge DOS is associated with a significant increase in $\Delta F$ at fixed $n_{\scalebox{0.7}{\textrm{2D}}}$, with the corresponding transparency points then shifted to shorter wavelengths in the strained QW. In the unstrained and compressively strained structures we respectively calculate quasi-Fermi level separations $\Delta F = 0.353$ and 0.397 eV at $n_{\scalebox{0.7}{\textrm{2D}}} = 10^{12}$ cm$^{-2}$ -- i.e.~a significant 44 meV increase in $\Delta F$, even at a relatively low (sub-threshold) injection.

Secondly, as described above, significantly more holes are available to contribute to the (TE-polarised) optical gain in the compressively strained QW, leading to significantly higher modal gain at fixed $n_{\scalebox{0.7}{\textrm{2D}}}$ than in the unstrained QW. At the highest considered carrier density $n_{\scalebox{0.7}{\textrm{2D}}} = 10^{13}$ cm$^{-2}$ we calculate a peak TE-polarised modal gain $\Gamma g = 19.8$ cm$^{-1}$ at $\epsilon_{xx} = -2$\%, which is approximately 75\% larger than the value $\Gamma g = 11.3$ cm$^{-1}$ calculated for the unstrained QW. In the unstrained QW we note that the calculated peak TM-polarised material gain 1069 cm$^{-1}$ at threshold is 59\% of the peak TE-polarised material gain, confirming that a significant fraction of the injected carriers in the unstrained QW are lost to the TM-polarised optical mode. This issue is mitigated at $\epsilon_{xx} = -2$\%, where we calculate that the peak TM-polarised material gain 39 cm$^{-1}$ is only 2\% of the peak TE-polarised material gain at threshold.


The calculated variation of the peak TE-polarised modal gain with $n_{\scalebox{0.7}{\textrm{2D}}}$ is shown in Fig.~\ref{fig:fixed_thickness_optical}(b), again for the $\epsilon_{xx} = 0$ (dashed red line) and $-2$\% (solid blue line) QWs. At fixed $n_{\scalebox{0.7}{\textrm{2D}}}$ the reduced VB edge DOS and associated increase in $\Delta F$ in the compressively strained QW leads to a significant enhancement of the modal gain. We note that the differential material gain at threshold $\frac{dg}{dn}$ is also significantly enhanced in the strained QW, as expected. \cite{Reilly_JQE_1994,Adams_JSTQE_2011}


For this series of QWs having fixed $\lambda$ and $t$, we calculate that $\Gamma$ remains effectively constant at 0.53\% as the $\epsilon_{xx}$ is varied from 0 to $-2$\%. On this basis we compute a threshold modal gain $\Gamma g_{\scalebox{0.7}{\textrm{th}}} = 9.5$ cm$^{-1}$, denoted by the horizontal dotted black line in Fig.~\ref{fig:fixed_thickness_optical}(b). The calculated threshold characteristics of this set of QWs --  $n_{\scalebox{0.7}{\textrm{th}}}$ and $\frac{dg}{dn}$ -- are shown in Fig.~\ref{fig:fixed_thickness_optical}(c). We note that $\frac{dg}{dn}$ is presented in three-dimensional units, in order to remove any explicit dependence on $t$. Examining the dependence of $n_{\scalebox{0.7}{\textrm{th}}}$ (closed red circles) on $\epsilon_{xx}$ we note a rapid reduction with increasing $\vert \epsilon_{xx} \vert$, with the calculated value $n_{\scalebox{0.7}{\textrm{th}}} = 2.6 \times 10^{12}$ cm$^{-2}$ at $\vert \epsilon_{xx} \vert = 2$\% being approximately one-third of the value $8.1 \times 10^{12}$ cm$^{-2}$ calculated for the unstrained QW. On the basis of the calculated trends in $n_{\scalebox{0.7}{\textrm{th}}}$ we conclude that there is significant benefit to increasing the compressive strain in the QW up to and beyond $\vert \epsilon_{xx} \vert = 1.5$\%. While the majority of the calculated reduction in $n_{\scalebox{0.7}{\textrm{th}}}$ occurs between $\vert \epsilon_{xx} \vert = 0$ and 1.5\%, for higher strains $\vert \epsilon_{xx} \vert \gtrsim 1.5$\% there is an additional, albeit smaller, calculated reduction in $n_{\scalebox{0.7}{\textrm{th}}}$: the calculated threshold carrier density of $3.1 \times 10^{12}$ cm$^{-2}$ at $\epsilon_{xx} = -1.5$\% is already close to that calculated at $\epsilon_{xx} = -2$\%.

We note that the calculated values of $\frac{dg}{dn}$ increase strongly with increasing $\vert \epsilon_{xx} \vert$, from $0.77 \times 10^{-16}$ cm$^{2}$ at $\epsilon_{xx} = 0$ to $2.41 \times 10^{-16}$ cm$^{2}$ at $\vert \epsilon_{xx} \vert = 2.0$\%, an increase by a factor of at least three. We note that (i) the calculated increase in $\frac{dg}{dn}$ is approximately linear in $\vert \epsilon_{xx} \vert$, and (ii) the calculated $\frac{dg}{dn}$ at $\vert \epsilon_{xx} \vert = 2.0$\% -- which can be further enhanced by reducing $t$, as we will discuss below -- is comparable to the value of $2.76 \times 10^{-16}$ cm$^{2}$ we have previously calculated for a type-I GaAs$_{0.87}$Bi$_{0.13}$/GaAs QW designed to emit at 1.55 $\mu$m. \cite{Broderick_JSTQE_2015} This suggests that the differential gain that can be achieved in an InP-based mid-infrared dilute bismide QW laser should be approximately equal to that achievable in a 1.55 $\mu$m GaAs-based dilute bismide device, with the characteristics of the latter predicted to exceed those of existing devices operating at $\lambda \approx 1$ $\mu$m (due to an improvement in performance associated with an increase in compressive strain in the QW). \cite{Broderick_JSTQE_2015}


Finally, we have computed the radiative current density at threshold $J_{\scalebox{0.7}{\textrm{rad}}}^{\scalebox{0.7}{\textrm{th}}}$ by integrating over the SE spectrum calculated at threshold ($n_{\scalebox{0.7}{\textrm{2D}}} = n_{\scalebox{0.7}{\textrm{th}}}$). We calculate that $J_{\scalebox{0.7}{\textrm{rad}}}^{\scalebox{0.7}{\textrm{th}}}$ decreases strongly with increasing strain -- from 48.8 to 28.7 A cm$^{-2}$ between $\vert \epsilon_{xx} \vert = 0$ and 1\% -- beyond which there is little further decrease, with $J_{\scalebox{0.7}{\textrm{rad}}}^{\scalebox{0.7}{\textrm{th}}} = 24.9$ A cm$^{-2}$ at $\vert \epsilon_{xx} \vert = 2$\%. Employing the Boltzmann approximation and writing $J_{\scalebox{0.7}{\textrm{rad}}}^{\scalebox{0.7}{\textrm{th}}} = e B n_{\scalebox{0.7}{\textrm{th}}}^{2}$ under the assumption of net charge neutrality, we have extracted the radiative recombination coefficient $B$. We calculate that $B$ increases approximately linearly with increasing $\vert \epsilon_{xx} \vert$, from $3.25 \times 10^{-12}$ cm$^{3}$ s$^{-1}$ in the unstrained QW to $1.61 \times 10^{-11}$ cm$^{3}$ s$^{-1}$ at $\vert \epsilon_{xx} \vert = 2$\%, a close to fivefold increase. The calculated values of $B$ here are again broadly comparable to those we have calculated previously for near-infrared GaAs-based dilute bismide QWs. \cite{Broderick_JSTQE_2015}


On the basis of this analysis we suggest that significant improvement in performance can be obtained by increasing the compressive strain up to $\vert \epsilon_{xx} \vert \approx 1.5$\%, with additional enhancement of $\frac{dg}{dn}$ achievable at higher $\vert \epsilon_{xx} \vert$. Overall, we therefore conclude that QWs having compressive strains $\vert \epsilon_{xx} \vert$ in the range 1.5 to 2.0\% should be sought to optimise overall performance. Our analysis then identifies a favourable combination of QW properties from the perspective of epitaxial growth. Recalling from Fig.~\ref{fig:structure_and_model}(c) that the Bi composition $y$ required to maintain fixed $\lambda$ decreases with increasing $\vert \epsilon_{xx} \vert$ (cf.~Fig.~\ref{fig:fixed_thickness_electronic}(a) for the set of fixed $t$ structures under consideration here), our calculations suggest that to obtain an optimised laser structure at given $\lambda$ using QWs of fixed $t$ entails minimising $y$. Given the difficulties associated with Bi incorporation during epitaxial growth -- in particular the general requirement to reduce growth temperature, which degrades material quality -- the optimised laser structures identified here are those which can be grown at the highest possible temperature for a given target $\lambda$. These structures can therefore be expected to also optimise the material quality achievable via established epitaxial growth, further enhancing the overall device performance. This recommendation to grow structures which simultaneously maximise $\vert \epsilon_{xx} \vert$ and minimise $y$ is likely to be restricted primarily by the intrinsic strain-thickness limitations of the In$_{x}$Ga$_{1-x}$As$_{1-y}$Bi$_{y}$ layer(s) in a given structure. However, on the basis of our estimated strain-thickness limits (cf.~Sec.~\ref{sec:InGaAsBi_band_structure}), we do not expect this to impede the growth of such structures.


\subsubsection{Fixed strain, variable thickness}
\label{sec:variable_thickness}


\begin{figure*}[t!]
	\includegraphics[width=1.00\textwidth]{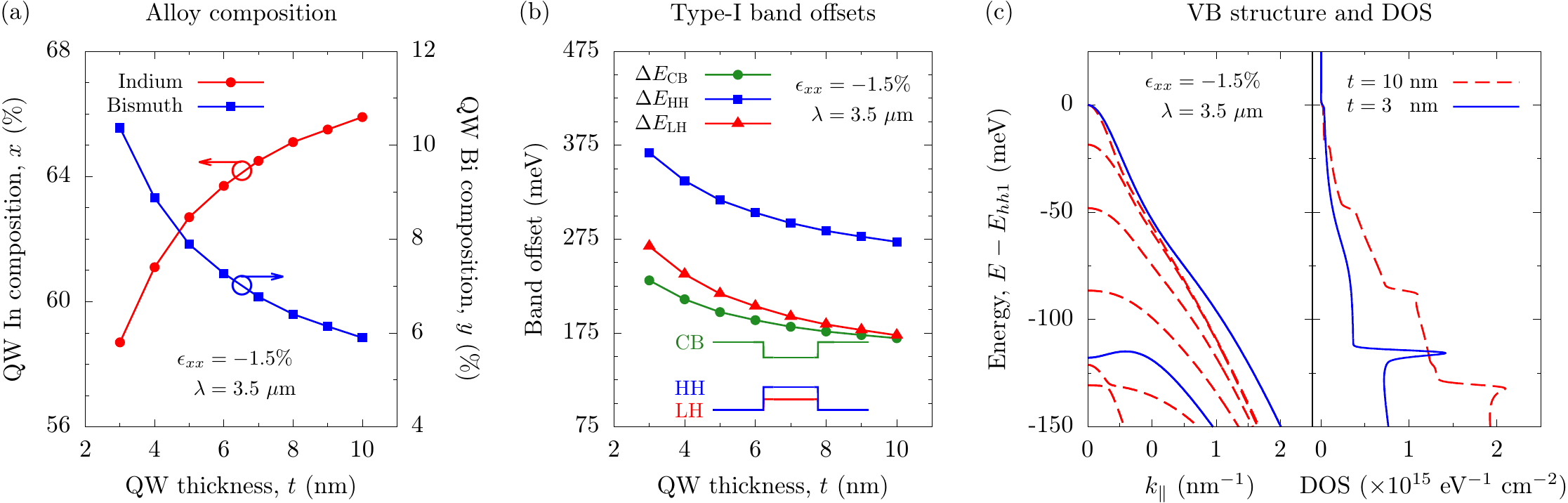}
	\caption{Summary of calculated electronic properties for In$_{x}$Ga$_{1-x}$As$_{1-y}$Bi$_{y}$ QWs having a fixed in-plane compressive strain $\epsilon_{xx} = -1.5$\%, variable thickness $t$, and emission wavelength $\lambda = 3.5$ $\mu$m at $T = 300$ K. (a) In$_{x}$Ga$_{1-x}$As$_{1-y}$Bi$_{y}$ QW alloy In and Bi compositions $x$ and $y$ required to maintain $\lambda = 3.5$ $\mu$m as a function of the QW thickness $t$. (b) Type-I CB (closed green circles), HH (closed blue circles) and LH (closed red triangles) band offsets, between the In$_{x}$Ga$_{1-x}$As$_{1-y}$Bi$_{y}$ QW and In$_{0.53}$Ga$_{0.47}$As barrier layers, as a function of $t$. (c) VB structure (left panel) and DOS (right panel) for the QWs having $t = 3$ nm (solid blue lines) and $t = 10$ nm (dashed red lines).}
     \label{fig:fixed_strain_electronic}
\end{figure*}

\begin{figure*}[t!]
	\includegraphics[width=1.00\textwidth]{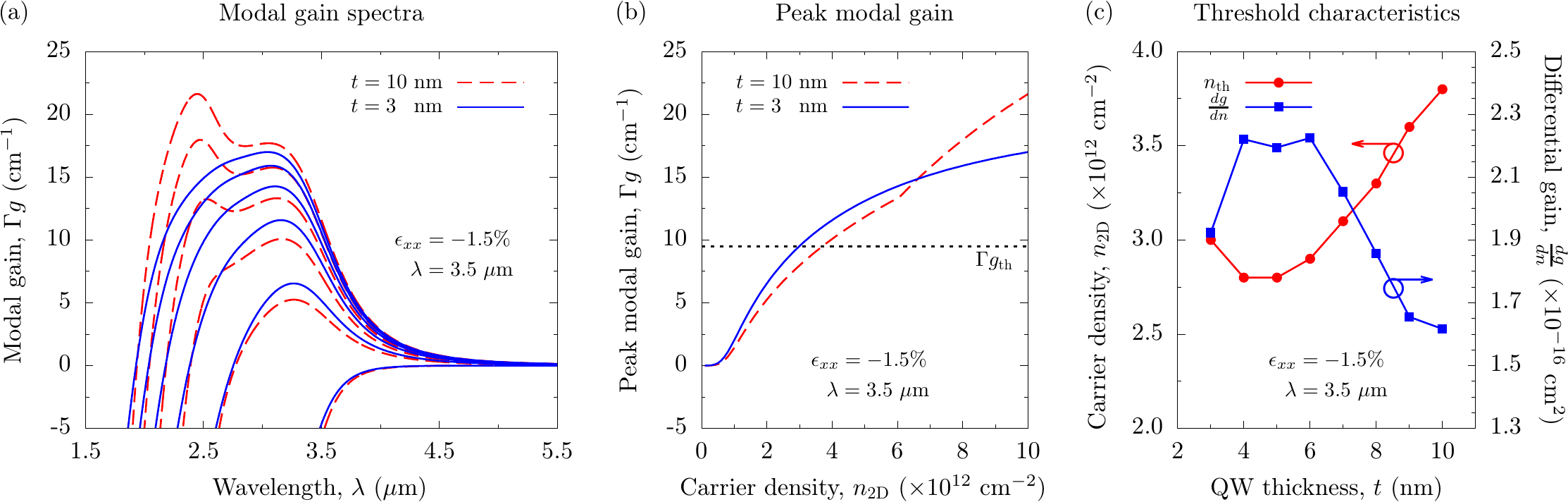}
	\caption{Summary of calculated optical properties for In$_{x}$Ga$_{1-x}$As$_{1-y}$Bi$_{y}$ QWs having a fixed in-plane compressive strain $\epsilon_{xx} = -1.5$\%, variable thickness $t$, and emission wavelength $\lambda = 3.5$ $\mu$m at $T = 300$ K. (a) TE-polarised modal gain spectra for the QWs having $t = 3$ nm (solid blue lines) and $t = 10$ nm (dashed red lines), calculated at sheet carrier densities $n_{\protect\scalebox{0.7}{\textrm{2D}}} = 10^{11}$ -- $10^{13}$ cm$^{-2}$ (in increments of $2 \times 10^{12}$ cm$^{-2}$). (b) Peak TE-polarised modal gain $\Gamma g$ as a function of $n_{\protect\scalebox{0.7}{\textrm{2D}}}$ for the same two QWs as in (a). The horizontal dotted black line denotes the calculated threshold modal gain, $\Gamma g_{\protect\scalebox{0.7}{\textrm{th}}} = 9.5$ cm$^{-1}$. (f) Threshold sheet carrier density ($n_{\protect\scalebox{0.7}{\textrm{th}}}$; closed red circles) and differential material gain at threshold ($\frac{dg}{dn}$; closed blue squares) as a function of $t$.}
     \label{fig:fixed_strain_optical}
\end{figure*}

The results of our calculations for $\lambda = 3.5$ $\mu$m laser structures having fixed $\epsilon_{xx}$ and variable $t$ are summarised in Figs.~\ref{fig:fixed_strain_electronic} and~\ref{fig:fixed_strain_optical}. Figure~\ref{fig:fixed_strain_electronic}(a) shows the calculated variation in $x$ (closed red circles) and $y$ (closed blue squares) required to maintain $\lambda = 3.5$ $\mu$m as $t$ is varied from 3 to 10 nm in QWs having fixed compressive strain $\epsilon_{xx} = -1.5$\%. At $t = 10$ nm we calculate that respective In and Bi compositions $x = 65.9$\% and $y = 5.9$\% are required to achieve $\lambda = 3.5$ $\mu$m with $\epsilon_{xx} = -1.5$\%. As $t$ is reduced the In (Bi) composition required to maintain fixed $\lambda$ and $\epsilon_{xx}$ decreases (increases), reaching $x = 58.7$\% ($y = 10.4$\%) at $t = 3$ nm. In this series of QWs the changes in $x$ and $y$ are required to compensate for the changes in confinement energy brought about by changes in $t$. As such, the changes in $x$ and $y$ required to maintain $\lambda = 3.5$ $\mu$m for this series of QWs are respectively significantly less than and approximately equal to those calculated for the fixed $t$ structures considered above. For the 10 nm thick QW we calculate a total confinement energy of 54 meV for the $e1$ and $hh1$ states, which increases approximately fourfold to 211 meV for a 3 nm thick QW. Recalling Fig.~\ref{fig:structure_and_model}(c) we note that these QWs, which have fixed compressive strain $\epsilon_{xx} = -1.5$\%, have alloy compositions which lie along a line in the composition space, running parallel to and equidistant between the $\epsilon_{xx} = -1.0$ and $-2.0$\% lines (dashed blue lines).


The calculated variation of the type-I CB, HH and LH band offsets for this series of QWs are shown in Fig.~\ref{fig:fixed_strain_electronic}(b) using, respectively, closed green circles, blue squares, and red triangles. In the thickest $t = 10$ nm In$_{0.659}$Ga$_{0.341}$As$_{0.941}$Bi$_{0.059}$ QW we calculate $\Delta E_{\scalebox{0.7}{\textrm{HH}}} = 272$ meV and $\Delta E_{\scalebox{0.7}{\textrm{LH}}} = 173$ meV. As the QW thickness is reduced we calculate that both $\Delta E_{\scalebox{0.7}{\textrm{HH}}}$ and $\Delta E_{\scalebox{0.7}{\textrm{LH}}}$ increase, reaching respective values of 367 and 267 meV in the 3 nm thick In$_{0.587}$Ga$_{0.413}$As$_{0.896}$Bi$_{0.104}$ QW. We note however that the difference between the HH and LH band offsets, which is determined by the axial component of the (fixed) strain in the QW layer, remains approximately constant at 100 meV irrespective of changes in $t$. We find that the calculated increase in $\Delta E_{\scalebox{0.7}{\textrm{HH}}}$ and $\Delta E_{\scalebox{0.7}{\textrm{HH}}}$ with decreasing $t$ is associated primarily with the increase in $y$ required to maintain $\lambda = 3.5$ $\mu$m, with the upward shift in the In$_{x}$Ga$_{1-x}$As$_{1-y}$Bi$_{y}$ HH and LH bulk band edge energies brought about by the increase in $y$ being sufficient to overcome the downward shift associated with the accompanying decrease in $x$. For the CB offset we calculate $\Delta E_{\scalebox{0.7}{\textrm{CB}}} = 170$ meV at $t = 10$ nm, increasing to 231 meV at $t = 3$ nm, with the downward shift in the In$_{x}$Ga$_{1-x}$As$_{1-y}$Bi$_{y}$ bulk CB edge energy brought about by the increase in $y$ sufficient to overcome the upward shift associated the accompanying decrease in $x$. We therefore conclude that the band offsets in QWs having fixed $\epsilon_{xx}$ are primarily determined by the Bi composition $y$, since the required $\lesssim 10$\% changes in the In composition $x$ required to maintain fixed $\lambda$ in response to changes in $t$ are too small to have a substantial impact on the electronic properties of the In$_{x}$Ga$_{1-x}$As host matrix semiconductor into which Bi is incorporated.


We again quantify the impact of the calculated band offsets on the expected laser performance by analysing (i) the localisation of the electron and hole charge density at $n_{\scalebox{0.7}{\textrm{th}}}$, (ii) the ionisation energies for bound electron and hole states, and (iii) the inter-band optical matrix elements at $k_{\parallel} = 0$ for $e1$-$hh1$ transitions. At threshold we calculate that 71.7\% (98.5\%) of the total electron (hole) charge density resides within the 10 nm thick QW, reducing to 30.9\% (93.7\%) as $t$ decreases to 3 nm, with the $t = 10$ and 3 nm QWs having respective threshold carrier densities $n_{\scalebox{0.7}{\textrm{th}}} = 3.8$ and $3.0 \times 10^{12}$ cm$^{-2}$. The calculated spill-out of the electron charge density in the narrow QW clearly indicates that electron confinement is significantly degraded as $t$ is reduced, despite an overall increase in $\Delta E_{\scalebox{0.7}{\textrm{CB}}}$. The calculated confinement energy for an $e1$ electron increases strongly with decreasing $t$, from 48 meV at $t = 10$ nm to 165 meV at $t = 3$ nm. As such, despite an overall increase in $\Delta E_{\scalebox{0.7}{\textrm{CB}}}$ with decreasing $t$, the ionisation energy for an $e1$ electron decreases strongly in narrower QWs, being roughly halved from 122 meV at $t = 10$ nm to 66 meV at $t = 3$ nm. Recalling that the ionisation energies described here, which are calculated at $k_{\parallel} = 0$, represent the maximum energy required for a bound electron to escape from the QW, it is clear that the filling of higher energy states contributes strongly -- as for the unstrained QW considered above -- to the thermal spill-out of the electron carrier density at threshold. As such, despite that the calculated ionisation energy in the 3 nm thick QW is $> 2 k_{\scalebox{0.7}{\textrm{B}}} T$ at room temperature, filling of higher energy electron states can be expected to contribute significantly to thermal leakage of electrons from the QW at and above room temperature. This expected increase in thermal leakage of electrons with decreasing $t$ is then an important consideration for the design of laser structures, suggesting that narrow QWs having $t \lesssim 5$ nm be avoided in order to avoid compromising efficiency and thermal stability. In accordance with this analysis of the electron confinement, we calculate respective $e1$-$hh1$ $k_{\parallel} = 0$ optical transition strengths of 14.59 and 8.98 eV in the $t = 10$ and 3 nm QWs, with the $\approx 40$\% decrease between $t = 10$ and 3 nm associated with (i) spill-out of the $e1$ envelope function from the narrow QW, as well as (ii) an increase in the Bi-localised character of the VB edge hole eigenstates due to the increased strength of the VBAC interaction accompanying the approximately 4\% increase in $y$ (cf.~Fig.~\ref{fig:fixed_strain_electronic}(a)).


The left- and right-hand panels of Fig.~\ref{fig:fixed_strain_electronic}(c) respectively show the calculated VB structure and DOS for the $t = 3$ nm (solid blue lines) and $t = 10$ nm (dashed red lines) QWs. For both QWs we have again chosen the zero of energy at the energy of the $k_{\parallel} = 0$ $hh1$ state, to facilitate direct comparison. For the 3 nm thick QW we calculate that there are four bound hole states: at $k_{\parallel} = 0$ the two highest energy states are HH-like ($hh1$ and $hh2$), while the third and fourth are respectively primarily LH- and HH-like ($lh1$ and $hh3$). Of the five highest energy bound hole states in the 10 nm thick QW, we calculate that the first four are HH-like at $k_{\parallel} = 0$ ($hh1$ -- $hh4$), while the fifth is primarily LH-like ($lh1$). We note that the compressive strain in both QWs leads to low in-plane hole effective masses close in energy to the QW VB edge, in line with our above analysis of QWs having fixed $t$ and variable $\epsilon_{xx}$. As $t$ decreases we note an approximately sixfold increase in the separation in energy between the two highest energy hole subbands -- $hh1$ and $hh2$ in the $\epsilon_{xx} = -1.5$\% QWs considered here, for all values of $t$ -- from 19 meV at $t = 10$ nm to 118 meV at $t = 3$ nm. Turning our attention to the VB DOS in the right-hand panel of Fig.~\ref{fig:fixed_strain_electronic}(c) we note that the increase in the energy separation between hole subbands with decreasing $t$ leads to a decrease in the DOS in the vicinity of the VB edge. We note however that this reduction only becomes pronounced for energies $\gtrsim 50$ meV below the QW VB edge, and is relatively minimal compared to the reduction calculated above for increasing $\vert \epsilon_{xx} \vert$ in QWs having fixed $t$.


We turn our attention now to the optical properties of this set of QWs, as summarised in Fig.~\ref{fig:fixed_strain_optical}. Figure~\ref{fig:fixed_strain_optical}(a) shows the TE-polarised modal gain spectrum for the $t = 3$ nm (solid blue lines) and $t = 10$ nm (dashed red lines) QWs, calculated for sheet carrier densities $n_{\scalebox{0.7}{\textrm{2D}}} = 0.1$, 2, 4, 6, 8 and $10 \times 10^{12}$ cm$^{-3}$. Recalling that (i) $g$ at fixed $n_{\scalebox{0.7}{\textrm{2D}}}$ is inversely proportional to $t$, and (ii) $\Gamma$ tends to increase roughly linearly with $t$, \cite{Kawano_book_2001,Chuang_book_2009} we note that the calculated modal gain $\Gamma g$ is largely insensitive to changes in $t$ in these fixed $\lambda$ structures. As such, the differences in the calculated $\Gamma g$ spectra for the QWs considered here arise primarily due to intrinsic differences in the electronic rather than structural properties of the QWs. Indeed, we note that both the $t = 3$ and 10 nm QWs have similar gain characteristics at low $n_{\scalebox{0.7}{\textrm{2D}}}$. At low $n_{\scalebox{0.7}{\textrm{2D}}}$ only hole states lying close in energy to the QW VB edge are occupied: the similarity in the VB structure and DOS of both QWs in the immediate vicinity of the VB edge (cf.~Fig.~\ref{fig:fixed_strain_electronic}(c)) produces similar modal gain, with the slightly lower DOS in the narrow QW giving rise to marginally higher gain. As $n_{\scalebox{0.7}{\textrm{2D}}}$ increases the large energy separation between the $hh1$ and $hh2$ subbands in the narrow QW leads to occupation of $hh1$ band states at large $k_{\parallel}$, while in the thick QW holes begin to occupy $hh2$ band states close to $k_{\parallel} = 0$. This filling of the available VB states brings about (i) slightly higher modal gain at fixed $n_{\scalebox{0.7}{\textrm{2D}}}$ in the narrow QW, due to more $hh1$ holes remaining available to recombine with $e1$ electrons, and (ii) the emergence of a second peak in the gain spectrum at higher energy (shorter wavelength) in the thick QW, which we identify as being related both to $e1$-$hh2$ and, at high $n_{\scalebox{0.7}{\textrm{2D}}}$, $e2$-$hh2$ transitions (occuring respectively for $k_{\parallel}$ between 0.5 and 1.0 nm$^{-1}$, and close to $k_{\parallel} = 0$).


We see these general trends borne out in the calculated variation of the peak TE-polarised modal gain with $n_{\scalebox{0.7}{\textrm{2D}}}$, shown in Fig.~\ref{fig:fixed_strain_optical}(b) using solid blue and dashed red lines for the 3 and 10 nm thick QWs respectively. We note on the basis of the discussion above that the calculated threshold modal gain for this series of structures varies little with changes in $t$, so that $\Gamma g_{\scalebox{0.7}{\textrm{th}}}$ remains approximately constant at 9.5 cm$^{-1}$ as in the fixed $t$ structures considered above. At low and intermediate $n_{\scalebox{0.7}{\textrm{2D}}}$ the modal gain in the narrow QW slightly exceeds that of the thick QW due primarily to the slightly lower DOS close in energy to the VB edge, with little difference in $\frac{dg}{dn}$ at threshold. At higher $n_{\scalebox{0.7}{\textrm{2D}}}$ the modal gain in the thick QW then exceeds that of the narrow QW, due to contributions from higher energy transitions associated with filling of $hh2$ VB states.


Figure~\ref{fig:fixed_strain_optical}(c) shows the calculated variation of the threshold characteristics with $t$ for this series of QWs. As $t$ is decreased from 10 nm, $n_{\scalebox{0.7}{\textrm{th}}}$ is reduced in line with the reduction in the VB DOS (brought about by the increased separation in energy between the $hh1$ and $hh2$ subbands). We determine that $n_{\scalebox{0.7}{\textrm{th}}}$ reaches a minimum value of $2.8 \times 10^{12}$ cm$^{-2}$ for a 4 -- 5 nm thick QW. As the QW thickness is reduced further $n_{\scalebox{0.7}{\textrm{th}}}$ again increases, due to the degradation in electron confinement in a narrower QW. The spill-out of the $e1$ envelope function at $t = 3$ nm leads, as described above, to a sharp decrease in the $e1$-$hh1$ optical transition strength which is sufficient to overcome any additional enhancement in gain due to a further reduction of the VB edge DOS. The calculated trend for the differential gain is similar: reducing $t$ initially brings about an increase in $\frac{dg}{dn}$ at threshold -- reaching a maximum value $\approx 2.2 \times 10^{-16}$ cm$^{2}$ for a 4 -- 6 nm thick QW -- but further decreases in $t$ bring about a degradation in the threshold characteristics of the device.


Using the SE spectra calculated at threshold we determine that the radiative current density $J_{\scalebox{0.7}{\textrm{rad}}}^{\scalebox{0.7}{\textrm{th}}}$ decreases with decreasing $t$,  by approximately 40\% from 36.1 A cm$^{-2}$ at $t = 10$ nm to 22.5 A cm$^{-2}$ at $t = 3$ nm. This corresponds to a calculated roughly threefold reduction in the radiative recombination coefficient, from $B = 1.56 \times 10^{-11}$ to $4.69 \times 10^{-12}$ cm$^{3}$ s$^{-1}$ as $t$ reduces from 10 to 3 nm, spanning roughly the same range as for the variable $\epsilon_{xx}$ QWs considered above.


On the basis of these calculations we conclude (i) that compressive strain plays the dominant role in determining the performance of In$_{x}$Ga$_{1-x}$As$_{1-y}$Bi$_{y}$ laser structures, but (ii) that minor additional improvements in performance can be brought about by the growth of narrow QWs. While QWs having thicknesses 4 nm $\lesssim t \lesssim$ 6 nm optimise the calculated threshold characteristics, we note that there is likely to be minimal benefit in practice to growing narrow QWs. Firstly, our calculations indicate that the large confinement energies associated with narrower QWs lead to significant spill-out of the electron charge density at and above room temperature, suggesting that thermal leakage of electrons is likely to limit device performance. Secondly, narrower QWs will be less robust against modifications to the electronic properties associated with QW thickness and composition fluctuations. Finally, narrower QWs require larger Bi compositions $y$ in order to achieve a given emission wavelength at fixed strain (cf.~Fig.~\ref{fig:fixed_strain_optical}), presenting further challenges from the perspective of epitaxial growth.

Overall, our analysis suggests that the optimal path towards realisation of In$_{x}$Ga$_{1-x}$As$_{1-y}$Bi$_{y}$ mid-infrared lasers is via the growth of structures having compressive strains $1.5\% \lesssim \vert \epsilon_{xx} \vert \lesssim 2.0$\%, and intermediate QW thicknesses 5 nm $\lesssim t \lesssim 7$ nm. Such structures should be well placed to manage the trade-offs between alloy composition, carrier confinement and low VB edge DOS, in such a manner as to deliver low threshold carrier densities $n_{\scalebox{0.7}{\textrm{th}}} \approx 3 \times 10^{12}$ cm$^{-2}$ per QW, as well as threshold values of differential material gain $\frac{dg}{dn} \approx 2 \times 10^{-16}$ cm$^{2}$ per QW, which should be roughly equivalent to those achievable in a 1.55 $\mu$m GaAs$_{1-x}$Bi/GaAs QW laser.


\begin{figure*}[t!]
	\includegraphics[width=1.00\textwidth]{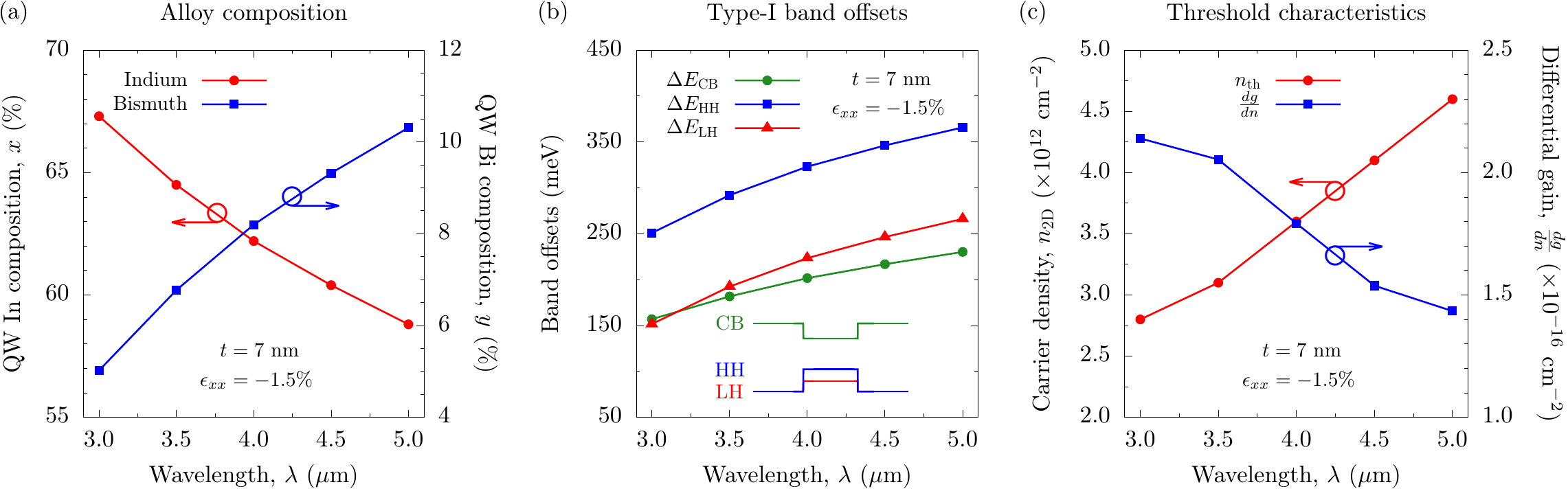}
	\caption{Summary of results for In$_{x}$Ga$_{1-x}$As$_{1-y}$Bi$_{y}$ QWs having fixed thickness $t = 7$ nm and in-plane compressive strain $\epsilon_{xx} = -1.5$\%, designed to have emission wavelengths $\lambda = 3$ -- 5 $\mu$m at $T = 300$ K. (a) In$_{x}$Ga$_{1-x}$As$_{1-y}$Bi$_{y}$ QW alloy In and Bi compositions $x$ and $y$ required to achieve $\lambda = 3$ -- 5 $\mu$m while maintaining fixed $t$ and $\epsilon_{xx}$. (b) Type-I CB (closed green circles), HH (closed blue circles) and LH (closed red triangles) band offsets, between the In$_{x}$Ga$_{1-x}$As$_{1-y}$Bi$_{y}$ QW and In$_{0.53}$Ga$_{0.47}$As barrier layers, as a function of $\lambda$. (c) Threshold sheet carrier density ($n_{\protect\scalebox{0.7}{\textrm{th}}}$; closed red circles) and differential material gain at threshold ($\frac{dg}{dn}$; closed blue squares) as a function of $\lambda$.}
     \label{fig:variable_wavelength_results}
\end{figure*}


\subsection{Dependence of laser performance on emission wavelength and number of quantum wells}
\label{sec:longer_wavelength}

Having elucidated trends in the predicted performance of 3.5 $\mu$m laser structures, we now turn our attention to the variation of laser properties with $\lambda$. We begin by noting that the qualitative trends in the performance identified at $\lambda = 3.5$ $\mu$m hold generally between 3 and 5 $\mu$m, and so the same general recommendations for QW design should be followed in order to optimise the device performance at a given emission wavelength. On the basis of our calculations above we focus here on QWs having thickness $t = 7$ nm and compressive strain $\epsilon_{xx} = -1.5$\%. To maximise $\Gamma$ at a given emission wavelength, it is necessary to carefully choose the thickness of the In$_{0.53}$Ga$_{0.47}$As barriers between the In$_{x}$Ga$_{1-x}$As$_{1-y}$Bi$_{y}$ QW and InP SCH layers of the structure. At $\lambda = 3.0$ $\mu$m we calculate that $\Gamma$ is maximised for 425 nm thick barriers both above and below the QW, and that this thickness should be increased by 25 nm for every subsequent 0.5 $\mu$m increase in $\lambda$ in order to maintain the maximum possible value of $\Gamma$. Despite this optimisation of the overall structure, we note that $\Gamma$ nonetheless decreases strongly with increasing wavelength. For example, for $t = 7$ nm we calculate $\Gamma = 0.63$\% in a structure designed to emit at 3 $\mu$m, which decreases to 0.37\% at $\lambda = 5$ $\mu$m. These low values of $\Gamma$ motivate the use of multi-QW structures so that $\Gamma$ can be maximised, and hence $n_{\scalebox{0.7}{\textrm{th}}}$ per QW reduced, in a given structure.


Our analysis of the dependence of the laser properties on $\lambda$ is summarised in Fig.~\ref{fig:variable_wavelength_results}, which presents a selection of our calculated results for QWs having $t = 7$ nm and $\epsilon_{xx} = -1.5$\%. Figure~\ref{fig:variable_wavelength_results}(a) presents the variation of the QW In and Bi compositions $x$ and $y$ -- shown using closed red circles and blue squares, respectively -- required to vary $\lambda$ between 3 and 5 $\mu$m for these fixed thickness, fixed strain structures. As expected, increasing the emission wavelength generally requires an increase in $y$, with $x$ then needing to be decreased accordingly to maintain constant $\epsilon_{xx}$. We determine that an In (Bi) composition $x = 67.3$\% ($y = 5.0$\%) is required to achieve emission at 3 $\mu$m, which must be decreased (increased) to $x = 58.8$\% ($y = 10.3$\%) to reach $\lambda = 5$ $\mu$m.

The variation of the CB, HH and LH band offsets with $\lambda$ for these structures is shown in Fig.~\ref{fig:variable_wavelength_results}(b) using, respectively, closed green circles, blue squares and red triangles. We predict that $\Delta E_{\scalebox{0.7}{\textrm{HH}}}$ and $\Delta E_{\scalebox{0.7}{\textrm{LH}}}$ should increase with increasing $\lambda$, as a consequence of the increase in $y$, with their difference remaining approximately constant due to the fixed compressive strain in the QW (cf.~Sec.~\ref{sec:3500_nm_structures}). We calculate $\Delta E_{\scalebox{0.7}{\textrm{HH}}} = 251$ meV ($\Delta E_{\scalebox{0.7}{\textrm{LH}}} = 152$ meV) at $\lambda = 3$ $\mu$m, which increases to $\Delta E_{\scalebox{0.7}{\textrm{HH}}} = 366$ meV ($\Delta E_{\scalebox{0.7}{\textrm{LH}}} = 266$ meV) at $\lambda = 5$ $\mu$m. Similarly, we calculate that $\Delta E_{\scalebox{0.7}{\textrm{CB}}}$ increases with increasing $\lambda$, from 157 meV at $\lambda = 3$ $\mu$m to 230 meV at $\lambda = 5$ $\mu$m. This increase in $\Delta E_{\scalebox{0.7}{\textrm{CB}}}$ is again determined by the increase in $y$, with the corresponding downward shift of the QW bulk CB edge energy sufficient to overcome the upward shift due to the reduction in $x$. We note that the corresponding ionisation energies for $e1$ electrons ($hh1$ holes) also increase with increasing $\lambda$, from 90 meV (238 meV) at $\lambda = 3$ $\mu$m to 141 meV (354 meV) at $\lambda = 5$ $\mu$m. Based on these trends, as well as our detailed analysis of carrier localisation at threshold in Sec.~\ref{sec:3500_nm_structures}, we conclude overall that carrier leakage at and above room temperature may be minimised in optimised In$_{x}$Ga$_{1-x}$As$_{1-y}$Bi$_{y}$ QWs, and hence should not limit the overall device performance at any wavelength $\gtrsim 3$ $\mu$m.

Figure~\ref{fig:variable_wavelength_results}(c) shows the variation of the calculated threshold characteristics -- $n_{\scalebox{0.7}{\textrm{th}}}$ and $\frac{dg}{dn}$ -- with $\lambda$ for this same set of QWs. We note an overall tendency here for the laser performance to degrade with increasing emission wavelength. The calculated threshold carrier density increases from $2.8 \times 10^{12}$ cm$^{-2}$ at $\lambda = 3$ $\mu$m to $4.6 \times 10^{12}$ cm$^{-2}$ at $\lambda = 5$ $\mu$m, a significant increase of close to two-thirds across the 3 -- 5 $\mu$m range. Likewise, we calculate a similar decrease in the differential gain at threshold, with the calculated value of $1.43 \times 10^{16}$ cm$^{-2}$ at $\lambda = 5$ $\mu$m being approximately two-thirds of the value $2.14 \times 10^{-16}$ cm$^{2}$ calculated at $\lambda = 3$ $\mu$m. The radiative current density at threshold $J_{\scalebox{0.7}{\textrm{rad}}}^{\scalebox{0.7}{\textrm{th}}}$ is calculated to increase from 27.4 to 33.1 A cm$^{-2}$ between $\lambda = 3$ and 5 $\mu$m. Bearing in mind the calculated increase in $n_{\scalebox{0.7}{\textrm{th}}}$ over this wavelength range, we note that this corresponds to an approximately 55\% decrease in the radiative recombination coefficient $B$, from $1.53 \times 10^{-11}$ cm$^{3}$ s$^{-1}$ at $\lambda = 3$ $\mu$m to $6.83 \times 10^{-12}$ cm$^{3}$ s$^{-1}$ at $\lambda = 5$ $\mu$m.


Since Bi incorporation primarily impacts the VB structure, leaving the CB structure relatively unperturbed, our analysis of the QW electronic properties suggests that trends in the CHCC Auger recombination rates should largely follow those observed in established devices operating in the same wavelength range. While a detailed investigation of the Auger recombination rates is beyond the scope of the present analysis, it is to be expected that CHCC Auger recombination -- the rate of which increases strongly with increasing $\lambda$ -- will be the dominant loss mechanism in the laser structures proposed here. \cite{Silver_JQE_1997} On the basis that our analysis has identified single QW structures which minimise $n_{\scalebox{0.7}{\textrm{th}}}$, and since the Auger contribution to the threshold current density is $\propto n_{\scalebox{0.7}{\textrm{th}}}^{3}$ (in the Boltzmann approximation, where $J_{\scalebox{0.7}{\textrm{Auger}}}^{\scalebox{0.7}{\textrm{th}}} = e C n_{\scalebox{0.7}{\textrm{th}}}^{3}$), we note that such optimised structures should also have minimised CHCC Auger losses. Specifically, structures designed to minimise $n_{\scalebox{0.7}{\textrm{th}}}$ can be expected to maximise the ratio $r = \frac{ J_{\scalebox{0.5}{\textrm{rad}}}^{\scalebox{0.5}{\textrm{th}}} }{ J_{\scalebox{0.5}{\textrm{Auger}}}^{\scalebox{0.5}{\textrm{th}}} } \sim \frac{ B n_{\scalebox{0.5}{\textrm{th}}}^{2} }{ C n_{\scalebox{0.5}{\textrm{th}}}^{3} }$ of the radiative and Auger contributions to the threshold current density thereby maximising the internal quantum efficiency $\eta_{i} = \frac{ J_{\scalebox{0.5}{\textrm{rad}}}^{\scalebox{0.5}{\textrm{th}}} }{ J_{\scalebox{0.5}{\textrm{th}}} } \approx \frac{ r }{ r + 1 }$ (where the approximation assumes minimal contributions to $J_{\scalebox{0.7}{\textrm{th}}}$ from monomolecular defect-related recombination and from carrier leakage, $J_{\scalebox{0.7}{\textrm{th}}} \approx J_{\scalebox{0.7}{\textrm{rad}}}^{\scalebox{0.7}{\textrm{th}}} + J_{\scalebox{0.7}{\textrm{Auger}}}^{\scalebox{0.7}{\textrm{th}}}$). \cite{Chuang_book_2009}


However, in light of (i) the aforementioned minor calculated increase in $J_{\scalebox{0.7}{\textrm{rad}}}^{\scalebox{0.7}{\textrm{th}}}$ between $\lambda = 3$ and 5 $\mu$m, and (ii) the expected strong increase in $J_{\scalebox{0.7}{\textrm{Auger}}}^{\scalebox{0.7}{\textrm{th}}}$ with increasing $\lambda$, it is to be expected that losses associated primarily with Auger recombination will act to reduce $r$, and hence $\eta_{i}$, with increasing $\lambda$. This degradation in performance can be partially mitigated by minimising $n_{\scalebox{0.7}{\textrm{th}}}$ per QW, thereby maximising $r$, suggesting that it may be beneficial in practice to utilise multi-QW structures to optimise performance.

In order to quantify the potential to further optimise overall performance by growing multiple QW structures, we have calculated the dependence of $n_{\scalebox{0.7}{\textrm{th}}}$ and $\frac{dg}{dn}$ on the number of QWs $N_{\scalebox{0.7}{\textrm{QW}}}$ for this series of structures having $t = 7$ nm and $\epsilon_{xx} = -1.5$\%, and $\lambda = 3$ -- 5 $\mu$m. At $\lambda = 3$ $\mu$m we calculate a significant reduction in $n_{\scalebox{0.7}{\textrm{th}}}$ per QW as $N_{\scalebox{0.7}{\textrm{QW}}}$ is increased from 1 to 3, from $2.8 \times 10^{12}$ cm$^{-2}$ in a single QW structure to $1.2 \times 10^{12}$ cm$^{-2}$ in a structure containing 3 QWs. Further increasing $N_{\scalebox{0.7}{\textrm{QW}}}$ results in only minor additional reductions in $n_{\scalebox{0.7}{\textrm{th}}}$ per QW, with the value of $1.1 \times 10^{12}$ cm$^{-2}$ calculated for a structure having $N_{\scalebox{0.7}{\textrm{QW}}} = 6$ representing the lower limit of the calculated trend. Our analysis therefore suggests that $n_{\scalebox{0.7}{\textrm{th}}}$ per QW can be well optimised at $\lambda = 3$ $\mu$m by growing structures containing 3 QWs. Further analysis reveals that this trend holds generally throughout the 3 -- 5 $\mu$m range. On the basis that minimising $n_{\scalebox{0.7}{\textrm{th}}}$ per QW can also be expected to maximise the ratio $r$ of the contributions to the threshold current density associated with radiative and Auger recombination -- thereby maximising the internal quantum efficiency $\eta_{i}$ -- we conclude that there is significant scope to minimise the overall threshold current density via the growth of multiple QW structures.

Turning our attention to the differential gain at threshold, we calculate that $\frac{dg}{dn}$ per QW increases significantly with increasing $N_{\scalebox{0.7}{\textrm{QW}}}$. At $\lambda = 3$ $\mu$m we calculate $\frac{dg}{dn} = 5.10 \times 10^{-16}$ cm$^{2}$ per QW for $N_{\scalebox{0.7}{\textrm{QW}}} = 3$, a greater than twofold increase over the value $2.14 \times 10^{-16}$ cm$^{2}$ calculated for a single QW structure. Contrary to the calculated trends in $n_{\scalebox{0.7}{\textrm{th}}}$ per QW, we find further increases in $\frac{dg}{dn}$ per QW for $N_{\scalebox{0.7}{\textrm{QW}}} > 3$ which, although sustained, become increasingly sublinear with increasing $N_{\scalebox{0.7}{\textrm{QW}}}$. Therefore, while the growth of triple QW structures can be expected to minimise the threshold carrier density, in practice it is likely to be beneficial to grow a small number of additional QWs in order to maximise the overall differential gain. Recalling from Sec.~\ref{sec:3500_nm_structures} that optimised structures are those having relatively large compressive strains $1.5\% \lesssim \vert \epsilon_{xx} \vert \lesssim 2.0$\%, the growth of multiple QW structures is likely to be limited in practice by structural degradation associated with accumulation of elastic energy (which will increase with increasing $N_{\scalebox{0.7}{\textrm{QW}}}$). As such, we recommend that optimised structures are those having $3 \lesssim N_{\scalebox{0.7}{\textrm{QW}}} \lesssim 5$.


On the basis of our analysis we conclude that optimised laser structures are those having (i) compressive strains $1.5\% \lesssim \vert \epsilon_{xx} \vert \lesssim 2.0$\%, (ii) QW thicknesses 5 nm $\lesssim t \lesssim 7$ nm, and (iii) containing between 3 and 5 QWs. Such structures can be expected to simultaneously minimise $n_{\scalebox{0.7}{\textrm{th}}}$ and maximise $\frac{dg}{dn}$, while ensuring high structural quality. We recall that all of the laser structures considered here have $\Delta_{\scalebox{0.7}{\textrm{SO}}} > E_{g}$ in the In$_{x}$Ga$_{1-x}$As$_{1-y}$Bi$_{y}$ QW layers. As such, we expect that IVBA and CHSH Auger recombination processes involving the SO VB to be reduced. Reduction of losses related to these processes is established in GaSb-based structures operating at wavelengths $\gtrsim 2.1$ $\mu$m, due to the larger values of $\Delta_{\scalebox{0.7}{\textrm{SO}}}$ in Sb-containing alloys. \cite{Cheetham_APL_2011,Bedford_JAP_2011} The dilute bismide structures identified here promise to bring these benefits to Sb-free, InP-based diode lasers for the first time. Furthermore, given the large type-I band offsets and electron and hole ionisation energies in these QWs, both of which tend to increase with increasing $\lambda$, we expect that thermal carrier leakage will be minimised in these devices. Given the importance of Auger recombination and thermal carrier leakage in limiting the performance of existing mid-infrared diode lasers, we therefore conclude that the InP-based In$_{x}$Ga$_{1-x}$As$_{1-y}$Bi$_{y}$ structures investigated here offer distinct potential to deliver performance exceeding that of existing devices, and that this can be achieved in a wavelength range beyond that currently accessible using existing InP-based structures.


\section{Conclusions}
\label{sec:conclusions}

We have presented a theoretical analysis and optimisation of the properties and performance of mid-infrared semiconductor lasers based on the dilute bismide alloy In$_{x}$Ga$_{1-x}$As$_{1-y}$Bi$_{y}$. Our calculations demonstrate that compressively strained In$_{x}$Ga$_{1-x}$As$_{1-y}$Bi$_{y}$ QWs constitute a promising approach to realising antimony-free laser diodes operating in the 3 -- 5 $\mu$m wavelength range. In particular, the ability to readily engineer the band structure to achieve favourable characteristics in QWs designed to emit between 3 and 5 $\mu$m and which can be grown on conventional (001) InP substrates promises to overcome several limitations associated with existing diode lasers operating in this challenging wavelength range. By considering SCH QW laser structures incorporating unstrained In$_{0.53}$Ga$_{0.47}$As barriers and InP cladding layers we demonstrated that it should be possible to achieve emission across the entire 3 -- 5 $\mu$m range.

Based on a theoretical model which we have previously used to quantitatively predict the optical gain in GaAs-based dilute bismide QW lasers, we have undertaken a comprehensive investigation of the electronic and optical properties of In$_{x}$Ga$_{1-x}$As$_{1-y}$Bi$_{y}$ laser structures designed to emit between 3 and 5 $\mu$m. The ability to independently select the strain and band gap (emission wavelength) in these quaternary QWs provides broad scope to engineer the band structure, enabling to achieve QWs with (i) large type-I band offsets, (ii) VB spin-orbit splitting energy in excess of the band gap, and (iii) low DOS in the vicinity of the VB edge. These properties allow for the development of lasers which, respectively, (i) suffer from minimal thermal leakage of carriers, (ii) are expected to exhibit reduced losses associated with IVBA and CHSH Auger recombination processes involving the SO VB, and (iii) possess low carrier threshold densities and high differential gain, which are comparable to those calculated previously for GaAs-based dilute bismide QW lasers designed to emit at 1.55 $\mu$m. On this basis, and based on (i) the fact that Bi incorporation strongly impacts the VB structure while leaving the CB relatively unperturbed, as well as (ii) the known dependence of Auger recombination processes on $\lambda$, it is possible that hot-electron producing CHCC Auger recombination could play an important role as an efficiency-limiting mechanism in In$_{x}$Ga$_{1-x}$As$_{1-y}$Bi$_{y}$ laser structures. Further investigations are required to identify and quantify the intrinsic loss mechanisms in these structures.

To design laser structures having optimal performance we carried out a systematic analysis in order to quantify the key trends determining the device properties, and their interplay, as functions of the QW alloy composition, structural properties, and emission wavelength. By considering structures having fixed QW thickness and variable strain, and structures having fixed strain and variable QW thickness, we identified that the calculated performance of a given In$_{x}$Ga$_{1-x}$As$_{1-y}$Bi$_{y}$ laser structure is primarily influenced by the compressive strain in the QW. This favours the growth of QWs which, at fixed emission wavelength, tend to have relatively higher In and lower Bi compositions. Our calculations suggest that optimum performance can be achieved using multiple QW structures containing between 3 and 5 QWs, with the individual QWs having compressive strains between 1.5 and 2\%, QW thicknesses between 5 and 7 nm. Through our analysis we have determined the precise alloy compositions required to achieve emission between 3 and 5 $\mu$m subject to these conditions, thereby specifying candidate prototypical structures for epitaxial growth and experimental investigation.

The simple laser structures considered here are readily compatible with established growth of bulk-like In$_{x}$Ga$_{1-x}$As$_{1-y}$Bi$_{y}$ epitaxial layers. Indeed, we expect that building upon established growth to develop QWs and laser structures in the manner suggested by our analysis offers a straightforward route to an initial practical demonstration of mid-infrared lasing in this emerging material system. The primary challenge offered by the laser structures proposed here from the perspective of epitaxial growth is the simultaneous requirement for incorporation of relatively large In and Bi compositions, $55\% \lesssim x \lesssim 70$\% and $5\% \lesssim y \lesssim 10$\%. However, given recent rapid developments in the establishment and refinement of the growth of dilute bismide alloys and heterostructures by a number of groups, we expect that this challenge can be overcome. Overall, we conclude that In$_{x}$Ga$_{1-x}$As$_{1-y}$Bi$_{y}$ alloys have significant potential for the development of Sb-free mid-infrared semiconductor diode lasers grown on conventional InP substrates, providing a new candidate class of heterostructures to overcome a number of the limitations associated with existing devices in the particularly challenging 3 -- 4 $\mu$m spectral range.


\section*{Acknowledgements}

The authors dedicate this work to the memory of Prof.~T.~Jeffrey C.~Hosea (University of Surrey, U.K.) and Prof.~Naci Balkan (University of Essex, U.K.), with whom they shared many enjoyable and fruitful years of collaboration. This work was supported by the Engineering and Physical Sciences Research Council, U.K. (EPSRC; project nos.~EP/K029665/1, EP/H005587/1 and EP/N021037/1), by Science Foundation Ireland (SFI; project no.~15/IA/3082), and by the National University of Ireland (NUI; via the Post-Doctoral Fellowship in the Sciences, held by C.A.B.). 



\end{document}